\documentclass[aps,prd,showpacs,twocolumn,superscriptaddress]{revtex4-1}

\usepackage{graphicx}
\usepackage{epsfig,graphics,subfigure,psfrag,amsmath,amssymb}
\usepackage{lineno}
\usepackage{dcolumn}
\usepackage{bm}
\usepackage{overpic}
\usepackage{xspace}
\usepackage{rotating}
\usepackage{color}
\usepackage{multirow,diagbox}
\usepackage{colortbl}
\usepackage{epstopdf}
\usepackage{lipsum}
\usepackage{mathtools}
\usepackage{cuted}
\usepackage[utf8]{inputenc}
\usepackage[colorlinks,linkcolor=blue,anchorcolor=blue,citecolor=blue,urlcolor=blue]{hyperref}
\usepackage[table]{xcolor}

\newcommand{\ra}{\rightarrow}
\newcommand{\eff}{\varepsilon}
\newcommand{\jpsi}{J/\psi}

\newcommand{\piz}{\pi^{0}}
\newcommand{\tripiz}{\pi^{0}\pi^{0}\pi^{0}}
\newcommand{\etap}{\eta^{\prime}}
\newcommand{\chisq}{\chi^{2}}

\newcommand{\ksks}{K_{S}^{0}K_{S}^{0}}

\newcommand{\gam}{\gamma}

\begin{document}
\title{\boldmath Study of $f_1(1420)$ and $\eta(1405)$ in the decay $\jpsi \to \gam \tripiz$}

\author{
M.~Ablikim$^{1}$, M.~N.~Achasov$^{4,c}$, P.~Adlarson$^{76}$, O.~Afedulidis$^{3}$, X.~C.~Ai$^{81}$, R.~Aliberti$^{35}$, A.~Amoroso$^{75A,75C}$, Q.~An$^{72,58,a}$, Y.~Bai$^{57}$, O.~Bakina$^{36}$
% ...
\\
\vspace{0.2cm}
(BESIII Collaboration)\\
\vspace{0.2cm} {\it
$^{1}$ Institute of High Energy Physics, Beijing 100049, People's Republic of China\\
% ...
$^{3}$ Bochum  Ruhr-University, D-44780 Bochum, Germany\\
$^{4}$ Budker Institute of Nuclear Physics SB RAS (BINP), Novosibirsk 630090, Russia\\
% ...
$^{35}$ Johannes Gutenberg University of Mainz, Johann-Joachim-Becher-Weg 45, D-55099 Mainz, Germany\\
$^{36}$ Joint Institute for Nuclear Research, 141980 Dubna, Moscow region, Russia\\
% ...
$^{57}$ Southeast University, Nanjing 211100, People's Republic of China\\
$^{58}$ State Key Laboratory of Particle Detection and Electronics, Beijing 100049, Hefei 230026, People's Republic of China\\
% ...
$^{72}$ University of Science and Technology of China, Hefei 230026, People's Republic of China\\
% ...
$^{75}$ University of Turin and INFN, (A)University of Turin, I-10125, Turin, Italy; (B)University of Eastern Piedmont, I-15121, Alessandria, Italy; (C)INFN, I-10125, Turin, Italy\\
$^{76}$ Uppsala University, Box 516, SE-75120 Uppsala, Sweden\\
% ...
$^{81}$ Zhengzhou University, Zhengzhou 450001, People's Republic of China\\
\vspace{0.2cm}
$^{a}$ Deceased\\
% ...
$^{c}$ Also at the Novosibirsk State University, Novosibirsk, 630090, Russia\\
% ...
}
}

\begin{abstract}
A partial-wave analysis is performed on the decay $J/\psi\to\gamma\tripiz$ within the $\tripiz{}$ invariant-mass region below 1.6~GeV$/c^{2}$, using $(10.09~\pm~0.04)\times10^{9} ~J/\psi$ events collected with the BESIII detector. 
Significant isospin-violating decays of $\eta(1405)$ and $f_1(1420)$ into $f_0(980)\piz$ are observed.
For the first time, three axial-vectors, $f_1(1285)$, $f_1(1420)$ and $f_1(1510)$, are observed to decay into $\tripiz{}$.
The product branching fractions of these resonances are reported.

\end{abstract}

\maketitle 

\section{Introduction}
The non-Abelian structure of quantum chromodynamics (QCD) predicts the existence of bound states beyond those in the constituent quark model, such as glueballs, which are formed from gluons~\cite{Amsler:2004ps,Klempt:2007cp_GlueballReview,Crede:2008vw}.
The identification of glueballs would provide further validation of the predictions of QCD and the study of glueballs thus plays an important role in the field of hadron physics.
However, the possible mixing of pure glueballs with nearby $q\bar{q}$ nonet mesons makes the identification of glueballs difficult, both experimentally and theoretically.
Glueballs are expected to be copiously produced in radiative $J/\psi$ decays~\cite{Klempt:2007cp_GlueballReview,Crede:2008vw,Sarantsev:2021ein,Rodas:2021tyb}, 
which are therefore regarded as an ideal hunting ground in the search of glueballs.

The first glueball candidate, $\iota(1440)$, was observed in radiative $\jpsi$ decay~\cite{Scharre:1980zh_markII,Edwards:1982nc_crystall}.
The $\iota(1440)$ is now generally considered to be formed from two states, $\eta(1405)$ and $\eta(1475)$~\cite{PDG:2022pth}, of which the lower-mass meson is still regarded as a potential glueball candidate despite its mass being significantly less than LQCD predictions~\cite{Bali:1993fb_glueballspectro,Morningstar:1999rf_glueballspectro,Chen:2005mg_glueballspectro}.
This is known as the long-standing ``E/$\iota$ puzzle''. 
Some commentators, however, consider the $\eta(1405)$ and $\eta(1475)$ to be a single state observed
in different decay modes~\cite{Klempt:2007cp_GlueballReview}. 
A further puzzle concerns the $\eta(1295)$, generally considered to be a radial excitation of the $\eta$ meson.  However, the existence of this state has been questioned\cite{Klempt:2006jk,Klempt:2007cp_GlueballReview}, though it has been observed in $\pi^{-}p$ experiments~\cite{E852:2001ote_eta1295ppi_1,Fukui:1991ps_eta1295ppi_2,GAMS:1997pxg_eta1295ppi_3,E852:2000rhq_eta1295ppi_4}, $p\bar{p}$ annihilation~\cite{eta1295_ppbar1_Abele:1998qd,eta1295_ppbar2_Anisovich:2001jb,eta1295_ppbar3_Amsler:2004rd}, and also seen in the decays
$\jpsi\to\gamma\eta\pi\pi$~\cite{eta1295_radjpsi_DM2:1990cwz} and 
$B\to\eta\pi\pi K$~\cite{eta1295_Bdecay_BaBar:2008rth}.  Clarifying this question has important consequences for the assignment of pseudoscalar glueball candidates.

The availability of a large sample of $\jpsi$ decays at BESIII provides the opportunity to learn more about the nature of the $\eta(1405)$ and $\eta(1475)$. 
BESIII has observed significant isospin-violating processes 
$\eta(1405)\ra f_{0}(980)\pi^{0}$ in the decay $\jpsi{}\ra\gam\eta(1405)\ra\gam\pi^{0}f_{0}(980)\ra\gam 3\pi$~\cite{isobreak_jpsitoG3pi_BESIII:2012aa}. 
The statistical significance of  the $\eta(1405)$ signal is found to be larger than 10$\sigma$ and 
an enhancement potentially from the 
$f_{1}(1285)/\eta(1295)$  is seen with a significance of 3.7$\sigma$ in the charged channel and with 1.2$\sigma$ in the neutral channel.
The width of the $f_{0}(980)$ measured in the $\pi\pi$ mass spectra is anomalously narrower than the world-average value. 
Interestingly, the isospin violation turns out to be significant, with ${\mathcal B}$($\eta(1405)\ra\pi^{0}f_{0}(980)\ra 3\pi$)/${\mathcal B}$($\eta(1405)\rightarrow\pi^{0}a_{0}(980)\rightarrow\eta\pi\pi$)=(17.9$\pm$4.2)\%~\cite{CrystalBarrel:1995kfe,PDG:2022pth,isobreak_jpsitoG3pi_BESIII:2012aa}, which cannot be explained by $a_{0}$-$f_{0}$ mixing. 
Based on this observation, a triangular singularity mechanism has been proposed~\cite{Wu:2011yx,Wu:2012pg,Du:2019idk,Cheng:2024sus} to explain the 
large isospin violation and the narrow line-shape of the $f_0(980)$.
The role of $f_1(1420)$ in the decay $\jpsi\ra\gam \tripiz{}$ has also been discussed in Ref.~\cite{Wu:2012pg}.

It has been proposed that the axial-vector meson $f_1(1285)$ is  a $K^{*}\bar{K}$ molecule~\cite{Aceti:2015pma}.
The LHCb result~\cite{LHCb:2013ged} rules out the tetraquark interpretation of $f_1(1285)$.
The measurement of the $f_1(1285)$ mixing angle between the strange and non-strange components of its wave function in the $q\bar{q}$ structure model~\cite{LHCb:2013ged} is also consistent with earlier determinations assuming that $f_1(1420)$ is another isoscalar in the $1^{++}$ nonet~\cite{Gidal:1987bn}.
The $f_1(1420)$ was observed in decays to $K^{*}\bar{K}$~\cite{CERN-CollegedeFrance-Madrid-Stockholm:1980umk} and $K\bar{K}\pi$~\cite{MARK-III:1990wgk}, and was proposed to be a hybrid meson~\cite{Ishida:1989xh}, a $K^{*}\bar{K}$ molecule~\cite{Longacre:1990uc}, or a manifestation of the $f_1(1285)$ in the $K^{*}\bar{K}$ decay~\cite{Debastiani:2016xgg}.
However, the absence of the $f_1(1420)$ in $K^{-}p$ reactions~\cite{Aston:1987ak} suggests that the less-established $f_1(1510)$~\cite{Close:1997nm}, rather than the $f_1(1420)$, may be the $s\bar{s}$ member of the $1^{++}$ nonet.
Further investigations into the decay properties of these three axial vector mesons will contribute to a deeper understanding of the nature of these states.

Recently, BESIII observed large contributions from $\eta(1405),~\eta(1475)$, $f_1(1285)$ and $f_1(1420)$ in the $\ksks{}\piz$ invariant-mass region of (1.1, 1.6) GeV/$c^{2}$ in $\jpsi\ra\gam\ksks{}\piz$ decays~\cite{BESIII:2022chl}, 
but no clear contribution from the $\eta(1295)$.
Because the processes $\eta(1405)/f_1(1420) \to f_0(980)\piz$ are isospin-violating in the final-state $\piz\piz\piz$, the possible backgrounds can be well suppressed.
To reveal the properties of pseudo-scalars and axial vectors around 1.3 GeV/$c^{2}$ and 1.4 GeV/$c^{2}$, we perform a partial-wave analysis (PWA) on $\jpsi\ra\gam \tripiz \, (\piz\ra\gam\gam)$ decays  in the $\tripiz{}$ invariant-mass ($M(\tripiz{})$)
region below 1.6 GeV/$c^{2}$, based on $(10.09~\pm~0.04)\times10^9~\jpsi$~\cite{numJpsi_BESIII:2021cxx} events collected at the center-of-mass energy of 3.097 GeV with BESIII detector.

\section{Detector and Monte Carlo simulations}

The BESIII detector~\cite{bes3detector_BESIII:2009fln} records symmetric $e^+e^-$ collisions provided by the BEPCII storage ring~\cite{bepc_Yu:2016cof} in the center-of-mass energy range from 1.84 to 4.95~GeV, with a peak luminosity of $1.1 \times 10^{33}~\text{cm}^{-2}\text{s}^{-1}$ achieved at $\sqrt{s} = 3.773~\text{GeV}$. 
The cylindrical core of the BESIII detector covers 93\% of the full solid angle and consists of a helium-based multilayer drift chamber~(MDC), a plastic scintillator time-of-flight system~(TOF), and a CsI(Tl) electromagnetic calorimeter~(EMC), which are all enclosed in a superconducting solenoidal magnet providing a 1.0~T magnetic field.
The magnetic field was 0.9~T in 2012, which affects 10.8\% of the total $J/\psi$ data.
The solenoid is supported by an octagonal flux-return yoke with resistive plate counter muon identification modules interleaved with steel. 
The charged-particle momentum resolution at $1~{\rm GeV}/c$ is $0.5\%$, and the ${\rm d}E/{\rm d}x$ resolution is $6\%$ for electrons from Bhabha scattering. 
The EMC measures photon energies with a resolution of $2.5\%$ ($5\%$) at $1$~GeV in the barrel (end-cap) region. 
The time resolution in the TOF barrel region is 68~ps, while that in the end-cap region was 110~ps.
The end-cap TOF system was upgraded in 2015 using multigap resistive plate chamber technology, providing a time resolution of 60~ps, which benefits 87\% of the data used in this analysis~\cite{tof1_li2017study,*tof2_2017study,*tof3_Cao:2020ibk}.

Simulated data samples produced with a {\sc geant4}-based~\cite{GEANT4:2002zbu} Monte Carlo (MC) package, which includes the geometric description of the BESIII detector and the detector response, are used to determine detection efficiencies and to estimate backgrounds. 
The simulation models the beam-energy spread and initial-state radiation in the $e^+e^-$ annihilations with the generator {\sc kkmc}~\cite{kkmc1_Jadach:2000ir,*kkmc2_Jadach:1999vf}.
The inclusive MC sample includes both the production of the $J/\psi$ resonance and the continuum processes incorporated in {\sc kkmc}~\cite{kkmc1_Jadach:2000ir,*kkmc2_Jadach:1999vf}.
All particle decays are modelled with {\sc evtgen}~\cite{evtgen1_Lange:2001uf,*evtgen2_Ping:2008zz} using branching fractions either taken from the Particle Data Group~(PDG)~\cite{PDG:2022pth}, when available, or otherwise estimated with {\sc lundcharm}~\cite{lund1_Chen:2000tv,*lund2_Yang:2014vra}.
Final-state radiation from charged final-state particles is incorporated using the {\sc photos} package~\cite{photos_Barberio:1990ms}.
The signal MC events for $\jpsi\ra\gam \piz\piz\piz$, with subsequent  $\piz\ra\gam\gam$ decays, are generated uniformly in phase-space (PHSP).

\section{Event selection}
In $\jpsi\ra\gam \tripiz{}$ decays, the final state consists of seven photons.
Photon candidates are identified using showers in the EMC. 
The deposited energy of each shower must be more than 25~MeV in the barrel region ($|\!\cos \theta|< 0.80$) and more than 50~MeV in the end-cap region ($0.86 <|\!\cos \theta|< 0.92$).
To suppress electronic noise and showers unrelated to the event, the difference between the EMC time and the time to the most energetic photon is required to be within $[-500,500]$~ns.
Candidate events are required to have no charged track.

Pairs of photon candidates forming the $\piz$ are selected with $\chi^2_{\rm 1C}<$ 10 by performing a one-constraint (1C) kinematic fit with the mass of each pair of photons constrained to the known mass of the $\piz$ meson~\cite{PDG:2022pth}.
The number of $\piz$ candidates is required to be at least three.
To reduce background events and improve mass resolution, a seven-constraint (7C) kinematic fit is performed under the hypothesis of $\jpsi\ra\gam\piz\piz\piz$, whose $\chi^{2}$ value is denoted as $\chi^{2}_{\rm 7C}(\gam\piz\piz\piz)$, imposing energy-momentum conservation (4C) and three extra  single $\piz$ constraints on each pair of photons (3C).
For events with more than one combination of $\jpsi\ra\gam \tripiz{}$, 
the combination with the lowest value of $\chi_{\rm 7C}^{2}(\gam\piz\piz\piz)$ is selected.
It is then required that $\chi_{\rm 7C}^{2}(\gam\piz\piz\piz)<$ 40.
To suppress background contributions with eight photons, 4C kinematic fits are performed separately under the hypotheses of $\jpsi\ra 7\gam$ and $\jpsi\ra8\gam$, which yield goodness-of-fit $\chi^{2}$ values of $\chi^{2}_{\rm 4C}(7\gam)$ and $\chi^{2}_{\rm 4C}(8\gam)$, respectively.
The condition $\chi^{2}_{\rm 4C}(7\gam)<\chi^{2}_{\rm 4C}(8\gam)$ is then required.  
A 7C kinematic fit is also performed under the hypothesis of $\jpsi\ra\gam\eta\piz\piz$, whose $\chi^{2}$ value is denoted as $\chi^{2}_{\rm 7C}(\gam\eta\piz\piz)$. 
To suppress backgrounds from $\jpsi{}\ra\gam\eta \piz\piz$, 
the condition $\chi^{2}_{\rm 7C}(\gam\piz\piz\piz)<\chi^{2}_{\rm 7C}(\gam\eta\piz\piz)$ is required.
To reject backgrounds related to $\gam\piz$ from $\omega$ decays, 
and also backgrounds of the mis-combined $\piz$ constructed
from radiative photon ($\gam_{r}$) and another photon from $\piz$ decays, the conditions $|M(\gam\pi^{0})-M(\omega)|>0.06$ GeV/$c^{2}$ and $|M(\gam_{r}\gam)-M(\pi^{0})|>0.02$ GeV/$c^{2}$ are required, respectively.
The $\gam_{r}$ is identified in a 7C kinematic fit.

\begin{figure}[htbp!]
	\includegraphics[width=0.5\textwidth]{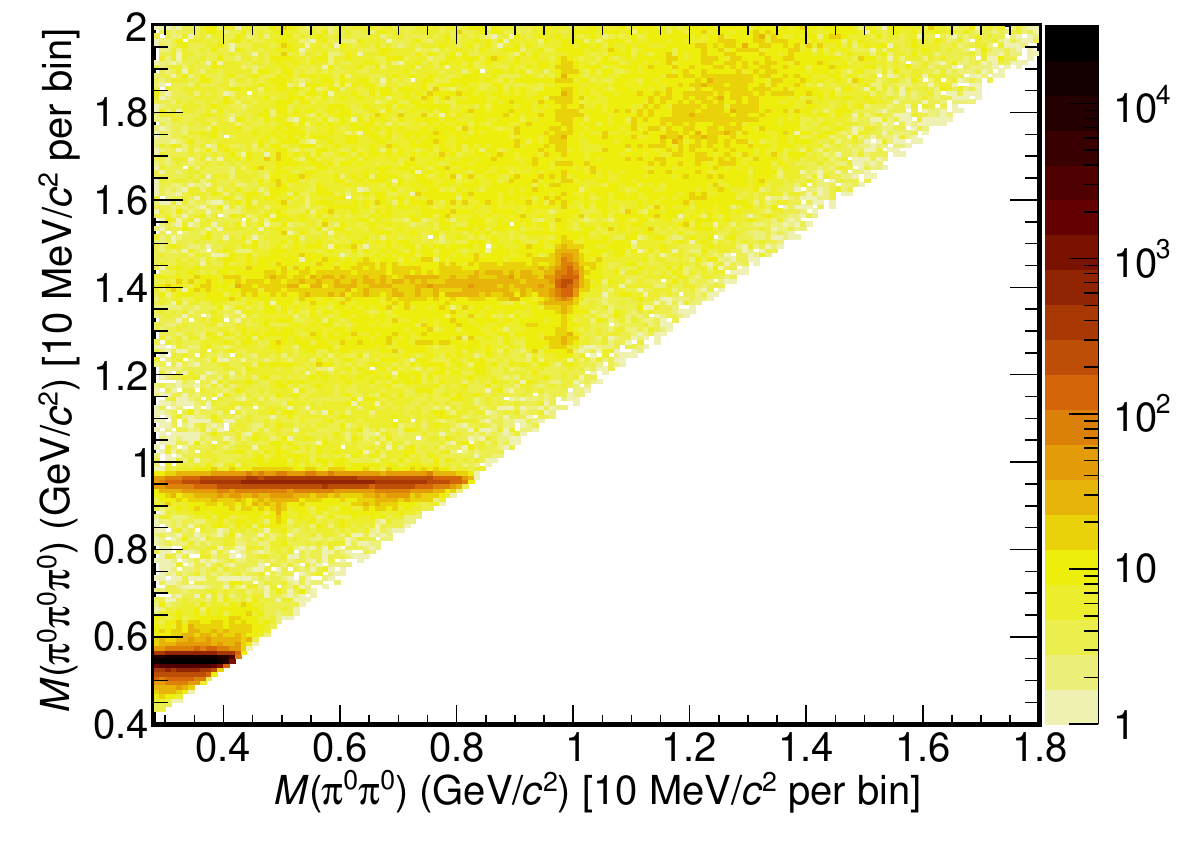}
	\caption{Distribution of $M(\piz\piz)$ versus $M(\tripiz{})$.
	\label{fig:sel_2pi_3pi}}
\end{figure}

\begin{figure*}[htbp!]
	\includegraphics[width=0.4\textwidth]{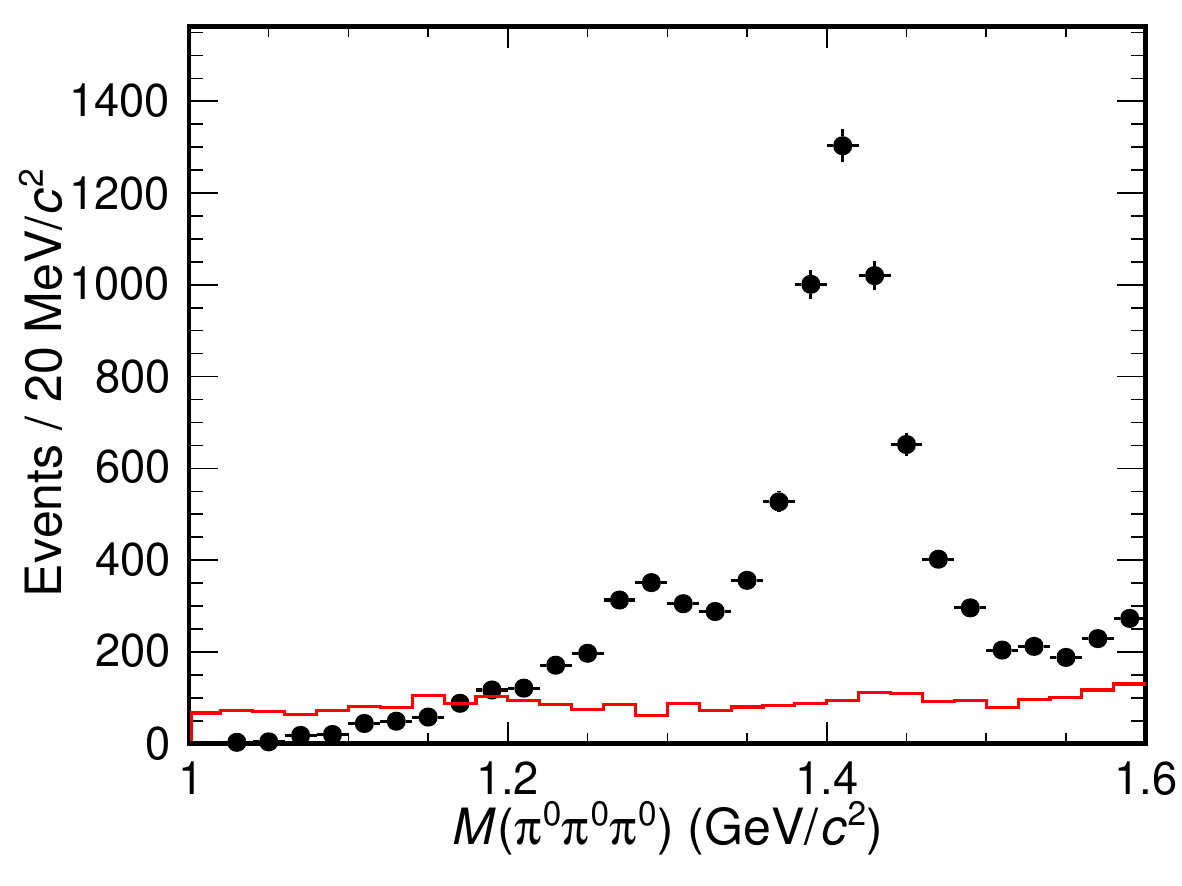} \put(-40,120){(a)}
	\includegraphics[width=0.4\textwidth]{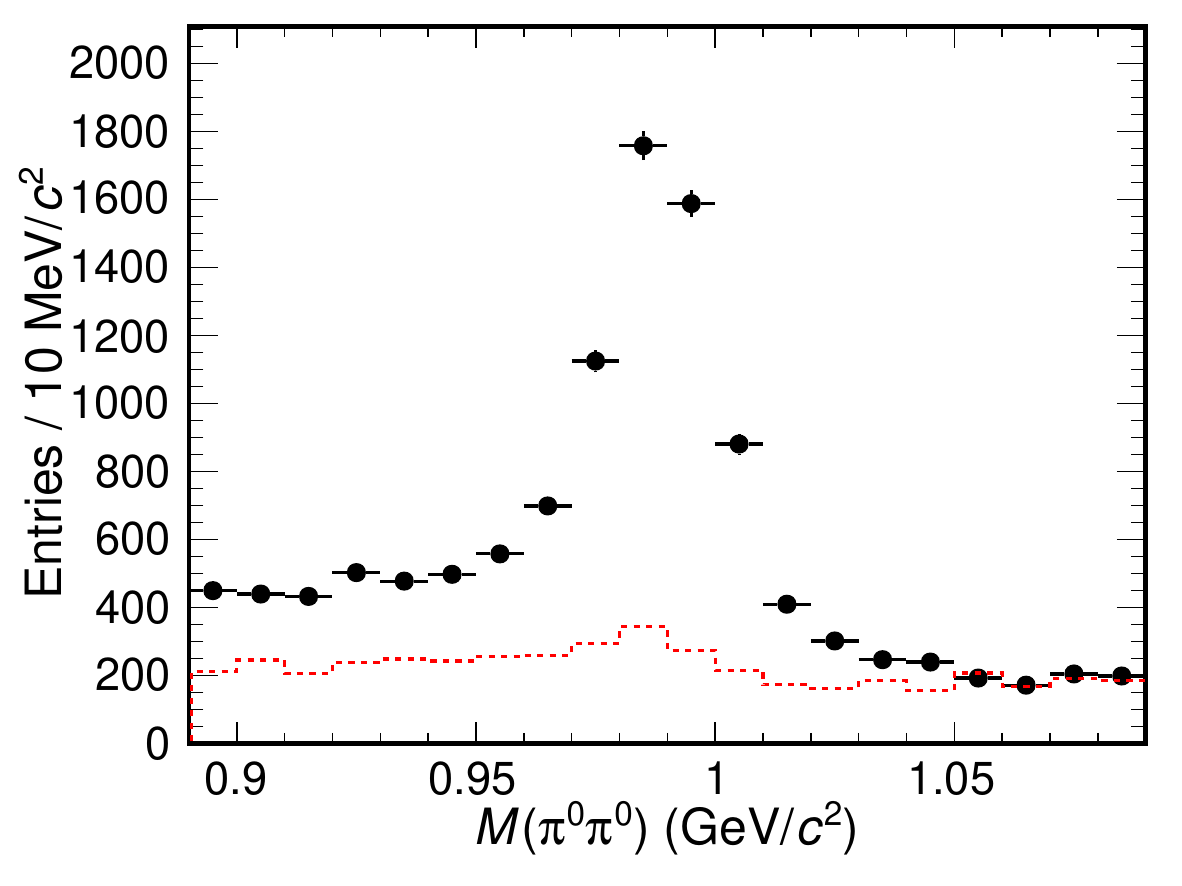} \put(-40,120){(b)} \\
	\includegraphics[width=0.4\textwidth]{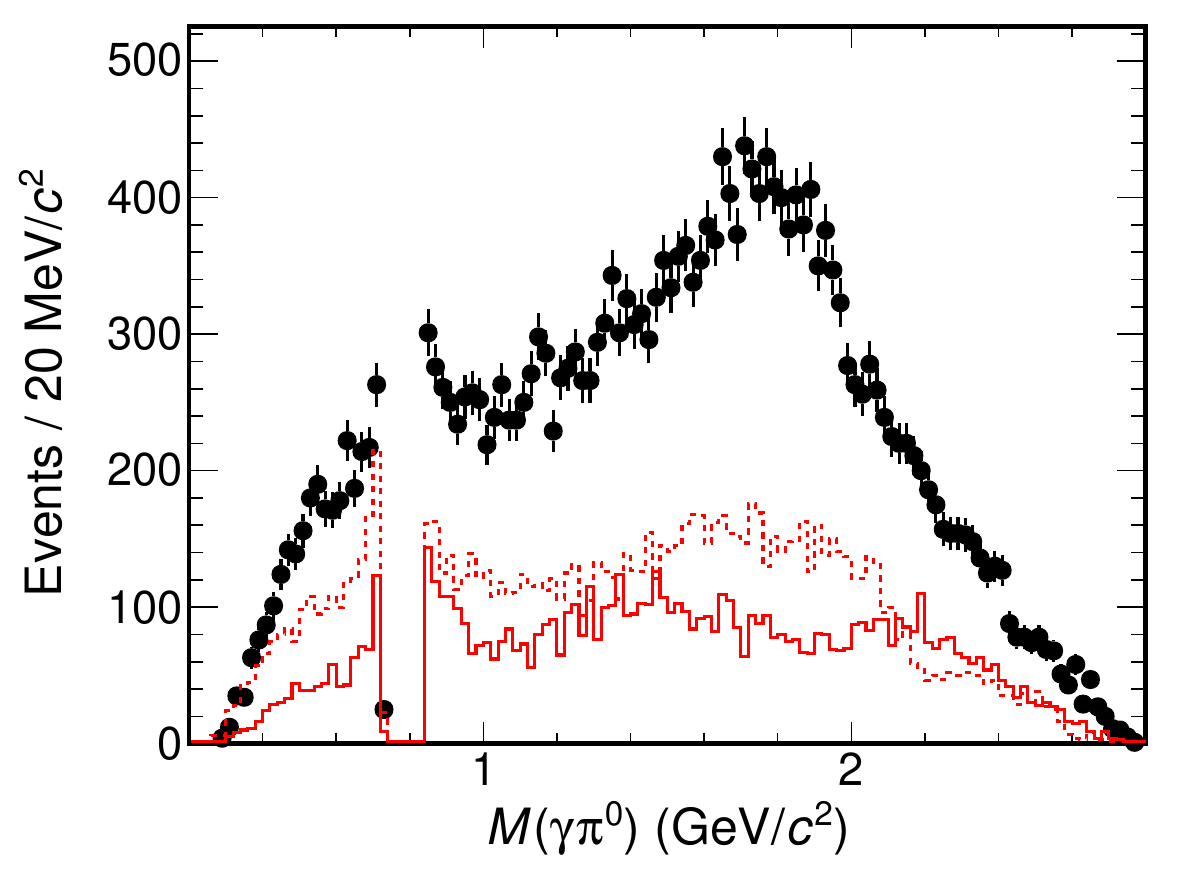} \put(-40,120){(c)}
	\caption{Distributions of (a) $M(\piz\piz\piz)$, (b) $M(\piz\piz)$ and (c) $M(\gam\piz)$,
	 where (b) and (c) are filled three times for each event.
	 The points with error bars show the data events used in PWA\label{fig:cut_3pi_2pi}; the red solid lines and dashed lines show the spectra from the sidebands of $f_0(980)$ and $\eta(1405)$, respectively.}
\end{figure*}

After imposing all the selection criteria above, a band around 1.4~GeV/$c^2$ in $M(\piz\piz\piz)$ that crosses with a band around 1.0~GeV/$c^2$ in $M(\piz\piz)$ can be seen clearly in the distribution of $M(\piz\piz\piz)$ versus $M(\piz\piz)$, as shown in Fig.~\ref{fig:sel_2pi_3pi}.
The other two horizontal bands that are visible are from $\etap$ and $\eta$ decays, respectively. 
The distributions of $M(\pi^{0}\pi^{0})$ from the $\eta(1405)$ sideband region,  defined by 0.20~GeV$/c^{2}\le\left|M(\pi^{0}\pi^{0}\pi^{0})-1.40\right|\le$ 0.40~GeV$/c^{2}$, 
and $M(\pi^{0}\pi^{0}\pi^{0})$ from the $f_{0}(980)$ sideband region, defined by 0.11~GeV$/c^{2}\le\left|M(\pi^{0}\pi^{0})-0.99\right|\le$ 0.21~GeV$/c^{2}$ from the closest $\piz\piz$ combination to the mass of $f_0(980)$, have also been checked. 
No significant peaks are observed in the spectrum of $M(\piz\piz\piz)$, while the $f_1(1285)$ and non-resonant process may have minor contribution to $f_0(980)$ in the spectrum of $M(\piz\piz)$, as shown by the red distributions in Fig.~\ref{fig:cut_3pi_2pi}.
Thus, $0^{++}$ PHSP is used to describe the non-resonant contribution in $M(\piz\piz)$ in the PWA.
To investigate the properties of pseudo-scalars and axial vectors around 1.3 and 1.4 GeV/$c^2$ in $M(\tripiz{})$, we require events that satisfy $M(\pi^{0}\pi^{0}\pi^{0})<1.6$ GeV/$c^{2}$.
As Fig.~\ref{fig:sel_2pi_3pi} indicates there is a clear concentration around 0.98 GeV/$c^{2}$ of $M(\pi^{0}\pi^{0})$, we require that each event should have at least a pair of $\pi^{0}$ in the $f_{0}(980)$ signal region within the mass window of [0.89, 1.09] GeV/$c^{2}$ to suppress possible backgrounds.
After applying these selections, 8810 events are retained.
The distributions of $M(\pi^{0}\pi^{0}\pi^{0})$ and $M(\pi^{0}\pi^{0})$ 
for the selected events are shown in Figs.~\ref{fig:cut_3pi_2pi}(a) and (b), respectively.
Structures are seen in Fig.~\ref{fig:cut_3pi_2pi}(a) around 1.3\,GeV/$c^2$ and 1.4\,GeV/$c^2$, which will be studied in the PWA fit.
A narrow peak of $f_0(980)$ can be seen clearly in $M(\pi^{0}\pi^{0})$ in Fig.~\ref{fig:cut_3pi_2pi}(b), which is consistent with the findings of a previous study~\cite{isobreak_jpsitoG3pi_BESIII:2012aa}. 

From a study using an inclusive MC sample of $10.01\times 10^{9}~\jpsi$ events 
with a generic event-type analysis tool~\cite{topo_Zhou:2020ksj}, the major backgrounds are found to be $\jpsi\ra\omega \piz\piz$ and $\jpsi\ra\gam\eta\piz\piz$.
The contributions of the backgrounds due to $\jpsi\ra\omega \piz\piz$ decays obtained from the MC in the distributions of $M(\piz\piz\piz)$ and $M(\piz\piz)$ are flat and can be described as non-resonant components 
in the PWA fit.
Because there are prominent intermediate resonances in the backgrounds from $\jpsi{}\ra\gam\eta\piz\piz$, a multi-dimensional re-weighting method~\cite{reweight_Liu:2018cad} is applied to the PHSP MC sample to obtain a ``data-like'' MC sample of $\jpsi{}\ra\gam\eta\piz\piz$ decays.
The weighted $\jpsi\ra\gam\eta\piz\piz$ MC events are subjected to the $\jpsi\ra\gam\piz\piz\piz$ event selection criteria.
The number of  events surviving is  
normalized according to the branching fraction and efficiency, resulting in 666$~\pm~230$ background events, which are subtracted in the PWA.

\section{Partial-wave analysis}
\subsection{Analysis method}
A PWA is performed to disentangle the structures in the $M(\piz\piz\piz)$ distribution 
using the GPUPWA framework~\cite{gpupwa_Berger:2010zza}.
The quasi-two-body decay amplitudes in the decay $\jpsi{}\ra\gam X$ with sequential decays $X\ra Y\piz,~Y\ra\piz\piz$ are constructed using the  covariant-tensor formalism described in Ref.~\cite{pwaform_Zou:2002ar}, where $X$ and $Y$ are the intermediate states.
Due to parity conservation and the absence of states with $J\ge2$ around the $f_0(980)$ mass, only the $0^{++}$ PHSP and $f_0(980)$ are considered for state $Y$.
Following Ref.~\cite{pwaform_Zou:2002ar}, for $\jpsi$ radiative decays to mesons including an intermediate resonance $X$, the covariant tensor amplitude $A_X$ is 
\begin{equation}
	A_X = \psi_{\mu}(m_1)e_{\nu}^{*}(m_2)A^{\mu\nu} = \psi_{\mu}(m_1)e_{\nu}^{*}(m_2)\sum_i\Lambda_{i}U_i^{\mu\nu},
\end{equation}
where $\psi_{\mu}(m_1)$ denotes the $\jpsi$ polarization four-vector with spin-projection $m_1$, $e_{\nu}^{*}(m_2)$ represents the polarization vector of the photon with spin projection $m_2$ and $U_i^{\mu\nu}$ stands for the $i$-th partial wave amplitude of $\jpsi$ radiative decays to intermediate resonance $X$ with a coupling strength determined by the complex parameter $\Lambda_i$.
The partial-wave amplitudes $U_i$ are constructed with the four-momenta of the particles in the final states, with specific expressions as given in Ref.~\cite{pwaform_Zou:2002ar}, where the exchange symmetry of identical particles has already been accounted for.

Each intermediate resonance is parametrized by a constant-width, relativistic Breit-Wigner (BW) propagator,
\begin{equation}
	BW(s) = \frac{1}{M^2-s-iM\Gamma},
\end{equation}
where $s$ is the square of $M(\tripiz{})$ or $M(\piz\piz)$, and 
$M$ and $\Gamma$ are the mass and width of the intermediate resonance, respectively.

The complex parameters of the amplitudes and resonance parameters, i.e. the masses and widths, are determined by an unbinned maximum likelihood fit.
The probability to observe the $i$-th event characterized by the measurement $\xi_i$, i.e. the measured four-momenta of the particles in the final states, is
\begin{equation}
	P(\xi_i) = \frac{\left|M(\xi_i)\right|^2\eff(\xi_i)\Phi(\xi_i)}{\sigma^\prime},
\end{equation}
where $\eff(\xi_i)$ is the detection efficiency, $\Phi(\xi_i)$ is the standard element of PHSP and $M(\xi_i) = \sum_X A_X(\xi_i)$ is the matrix element describing the radiative $\jpsi$ decay to the final state 
$\tripiz{}$ via intermediate resonance $X$.
The denominator $\sigma^\prime \equiv \int \text{d}\xi \left|M(\xi)\right|^2\eff(\xi) \Phi(\xi)$ is the normalization integral.

Due to the narrow width of $f_0(980)$ in the $M(\piz\piz)$ distribution,
an approximate numerical convolution of BW with a Gaussian function is applied to take into account the resolution ($9.6\pm0.3$ MeV). 
The $\mu$ and $\sigma$ of the Gaussian function 
and the resonance parameters of $f_0(980)$
are obtained by fitting the $M(\piz\piz)$ 
distribution and are subsequently fixed in the PWA.

The likelihood for observing $N$ events in the data sample is 
\begin{equation}
	{\cal L} = \prod^N_{i=1}P(\xi_i)=\prod^N_{i=1}\frac{\left|M(\xi_i)\right|^2\eff(\xi_i)\Phi(\xi_i)}{\sigma^\prime},
\end{equation}
and the fit for a given data set minimizes $-{\rm ln}~{\cal L}$, which is 
\begin{equation}
	\begin{aligned}
	-{\rm ln}~{\cal L} & =  -\sum^N_{i=1}{\rm ln}\left(\frac{\left|M(\xi_i)\right|^2}{\sigma^\prime}\right)-\sum^N_{i=1}{\rm ln}~\left(\eff(\xi_i)\Phi(\xi_i)\right) \\
	& = -\sum^N_{i=1}{\rm ln}\left|M(\xi_i)\right|^2+ N~{\rm ln}~\sigma^\prime -\sum^N_{i=1}{\rm ln}~\left(\eff(\xi_i)\Phi(\xi_i)\right),
	\end{aligned}
\end{equation}
In the fit the third term is ignored, as it  is constant and has no impact on the determination of the parameters or on the related changes of ${\rm -ln}~{\cal L}$.

The free parameters are optimized by using MINUIT~\cite{minuit_James:1975dr}.
The normalization integral $\sigma^\prime$ is evaluated using MC techniques.
An MC sample of $N_\text{gen}$ events uniformly distributed in PHSP is generated.
These events are subjected to the same selection criteria applied to the data and yield a sample of $N_\text{acc}$ accepted events.
The normalization integral is computed as 
\begin{equation}
	\sigma^\prime = \int \text{d}\xi \left|M(\xi)\right|^2 \eff(\xi) \Phi(\xi)  \propto \frac{1}{N_\text{gen}}\sum_k^{N_\text{acc}}\left|M(\xi_k)\right|^2.
\end{equation}
To take into account the $\eta\piz\piz$ background 
contribution in data, the negative log-likelihood (NLL) value obtained from re-weighted $\eta\piz\piz$ MC events, $-\sum_i\omega_i\cdot{\rm ln}~{\cal L}_{\text{bkg}_i}$ is subtracted from $-{\rm ln}~{\cal L}_\text{data}$, i.e., 
\begin{equation}
	-{\rm ln}~{\cal L} = -\alpha\left({\rm ln}~{\cal L}_\text{data} - \sum_i\omega_{{\rm bkg}_i}\cdot{\rm ln}~{\cal L}_{\text{bkg}_i} \right),
\end{equation}
where $\omega_{{\rm bkg}_i}$ represents the scaling factor of each background event and $\alpha$ is the normalization factor derived from Ref.~\cite{Langenbruch:2019nwe} to achieve an unbiased uncertainty estimation, which can be expressed as
\begin{equation}
	\alpha = \frac{\sum_{i\in {\rm data \& bkg}}~\omega_i}{\sum_{i \in{\rm data \& bkg}} \omega_i^2}.
\end{equation}

The number of fitted events $N_X$ for an intermediate resonance $X$ is defined as 
\begin{equation}
	N_X= \frac{\sigma_X}{\sigma^\prime}\cdot N^\prime,
\end{equation}
where $N^\prime$ is the number of selected events after background subtraction and 
\begin{equation}
	\sigma_X = \frac{1}{N_\text{gen}}\sum_k^{N_\text{acc}}\left| A_X(\xi_k) \right|^2
\end{equation}
is calculated with the same MC sample as the normalization integral $\sigma^\prime$.
Similarly, the number of fitted interference between an intermediate resonance $X_1$ and another intermediate resonance $X_2$ is defined as 
\begin{equation}
	N_{X_1,X_2}= \frac{\sigma_{X_1,X_2}}{\sigma^\prime}\cdot N^\prime,
\end{equation}
where
\begin{equation}
	\sigma_{X_1,X_2} = \frac{1}{N_\text{gen}}\sum_k^{N_\text{acc}}2Re[A_{X_1}(\xi_k)A_{X_2}(\xi_k)^*]. 
\end{equation}
The ratios $\sigma_X\slash\sigma^\prime$ and $\sigma_{X_1,X_2}\slash\sigma^\prime$ are then the fitted fractions for the intermediate resonance $X$ and the interference between an intermediate resonance $X_1$ and another intermediate resonance $X_2$, respectively. 
The detection efficiency $\eff_X$ for an intermediate resonance $X$ is obtained from the partial-wave amplitude weighted MC sample,
\begin{equation}
	\eff_X = \frac{\sum_k^{N_\text{acc}}\left| A_X(\xi_k)\right|^2}{\sum_i^{N_\text{gen}}\left| A_X(\xi_i)\right|^2}.
\end{equation}
The branching fraction of $\jpsi\ra\gam X,~X\ra \piz f_0(980)/0^{++}~{\rm PHSP} \ra \piz\piz\piz$ is  
\begin{equation}
\label{eq:br}
	{\cal B}(\jpsi\ra\gam X \ra\gam \piz\piz\piz)=\frac{N_{X}}{N_{\jpsi}\cdot {\cal B}_{\piz\ra\gam\gam}^{3}\cdot \eff_{X}}
	,
\end{equation}
where $N_{\jpsi}$ is the total number of $\jpsi$ events, and ${\cal B}_{\piz\ra\gam\gam}$ is the branching fraction of $\piz\ra\gam\gam$ quoted from Ref.~\cite{PDG:2022pth}.

\subsection{PWA results} \label{sub:pwa_result}
In this analysis, all combinations of possible resonances in the PDG~\cite{PDG:2022pth} with $J^{PC}=0^{-+},~1^{++}$ in $\piz\piz\piz$, including $\eta(1405)$, $f_1(1285)$, $f_1(1420)$, $f_1(1510)$, $\eta(1295)$, $\eta(1475)$, are considered. 
Possible non-resonant contributions in the $\tripiz{}$ system are described by $0^{-+}$ PHSP and $1^{++}$ PHSP.
In the $\piz\piz$ system, we only consider the contributions from $f_0(980)$ and $0^{++}$ PHSP.
The distinct narrow peak observed in the $\piz\piz$ invariant-mass spectrum related to resonances on $3\piz$ is much larger than the non-resonant contributions, which can be explained by dynamic mechanisms near the  $K^{*}K$ thresholds~\cite{Wu:2011yx,Wu:2012pg,Du:2019idk,Cheng:2024sus}.
Therefore, the interference between the processes with non-resonant contribution both on $\tripiz$ and $\piz\piz$, 
i.e. $0^{-+}~{\rm PHSP}(\tripiz)\to \piz + 0^{++}~{\rm PHSP}(\piz\piz)$ and $1^{++}~{\rm PHSP}(\tripiz)\to \piz + 0^{++}~{\rm PHSP}(\piz\piz)$, 
and other processes with resonances are ignored in the PWA fit.

Changes in the NLL value and the number of free parameters in the fits with and without a resonance included 
are used to evaluate the statistical significance of each component. 
All components are retained in the baseline model, except for the $\eta(1295)$ and $\eta(1475)$, for reasons discussed later.  

The baseline model includes $\eta(1405)$, $f_1(1285),~f_1(1420),~f_1(1510)$, and the non-resonant decay $\jpsi\ra\gam\piz\piz\piz$, which is modeled by $0^{-+}$ PHSP and $1^{++}$ PHSP of the $\piz\piz\piz$ system, as listed in Table~\ref{tab:base_solu}. 
Due to the limited sample size, only the masses and widths of $\eta(1405)$ and $f_1(1420)$ are free parameters in the PWA fit, while the masses and widths of the other resonances are fixed to the PDG values~\cite{PDG:2022pth}. 
The masses and widths of all resonances in the baseline model, the product branching fractions of $\jpsi\ra\gam X,~X\ra\piz f_0(980)/0^{++}~{\rm PHSP} \ra\tripiz{}$, and the statistical significances are summarized in Table~\ref{tab:base_solu}, 
where the first uncertainties are statistical and the second systematic.
The fit fractions of each component and their interference fractions are listed in Table~\ref{tab:frac_fit}.
\begin{table*}[htbp!] 
\renewcommand\arraystretch{1.5}
	\begin{center}
		{
		\caption{Masses, widths, {$\cal B$}($\jpsi\ra\gam X\ra\gam \piz f_0(980)/0^{++}{\rm PHSP} \ra\gam \tripiz{}$) and significance of each component in the baseline model, where the first uncertainties are statistical and the second 
		systematic.}
		\label{tab:base_solu}
		}
		\begin{tabular}{lcccc}
		\hline \hline		
		Resonance & $M$~(MeV/$c^2$) & $\Gamma$~(MeV) & ${\mathcal B}$ & Significance~($\sigma$) \\ \hline	
		$\eta(1405)$ & $1404^{+0.8}_{-1.5}{}^{+2.0}_{-8.1}$ & $46^{+1.8}_{-2.0}{}^{+4.2}_{-0.0}$ & (4.62$\pm$0.15$^{+5.08}_{-0.18}$)$\times 10^{-6}$ & $19.1$ \\
		$0^{-+}$ PHSP & --- & --- & (3.24$\pm$0.08$^{+0.41}_{-1.54}$)$\times 10^{-5}$ & $24.8$ \\
		$f_1(1285)$ & 1281.9 & 22.7 & (5.64$\pm$0.45$^{+0.74}_{-3.05}$)$\times 10^{-7}$ & $13.3$ \\
		$f_1(1420)$ & $1418^{+1.7}_{-2.1}{}^{+2.0}_{-2.2}$ & $46^{+3.4}_{-2.3}{}^{+6.1}_{-11.0}$ & (2.23$\pm$0.16$^{+0.20}_{-1.20}$)$\times 10^{-6}$ & $13.7$ \\
		$f_1(1510)$ & 1518 & 73 & (7.91$\pm$1.20$^{+0.74}_{-3.83}$)$\times 10^{-7}$ & $8.8$ \\
		$1^{++}$ PHSP & --- & --- & (2.60$\pm$0.08$^{+1.48}_{-1.66}$)$\times 10^{-5}$ & $13.3$ \\	
		\hline \hline		
		\end{tabular}
	\end{center}
\end{table*}

\begin{table*}[htbp!]
\renewcommand\arraystretch{1.5}

	\begin{center}
		{
		\caption{Fractions of each component and interference fractions between two components (\%) in the baseline model, where the uncertainties are statistical only.}
		\label{tab:frac_fit}
		}
		\begin{tabular}{lcccccc}
		\hline \hline
    Resonance & $\eta(1405)$ & $0^{-+}$ PHSP & $f_1(1285)$ & $f_1(1420)$ & $f_1(1510)$ & $1^{++}$ PHSP \\ \hline
    $\eta(1405)$ & 44.4$\pm$1.3 & -8.0$\pm$0.4 & 0.8$\pm$0.1 & 0.0$\pm$0.1 & -0.2$\pm$0.1 & -0.9$\pm$0.0 \\ 
    $0^{-+}$ PHSP  & --- & 25.6$\pm$0.5 & 0.0$\pm$0.0 & -0.1$\pm$0.0 & -0.0$\pm$0.0 & 1.0$\pm$0.1 \\ 
    $f_1(1285)$  & --- & --- & 4.8$\pm$0.4 & -1.8$\pm$0.4 & 1.4$\pm$0.2 & -0.8$\pm$0.2 \\ 
    $f_1(1420)$  & --- & --- & --- & 23.7$\pm$1.5 & -10.9$\pm$1.0 & 4.0$\pm$0.7 \\ 
    $f_1(1510)$  & --- & --- & --- & --- & 6.5$\pm$0.8 & -4.0$\pm$0.5 \\ 
    $1^{++}$ PHSP  & --- & --- & --- & --- & --- & 14.3$\pm$0.4 \\ 
		\hline \hline		
		\end{tabular}
	\end{center}
\end{table*}

\begin{figure*}[htbp!]
	\centering
	\includegraphics[width=0.38\textwidth]{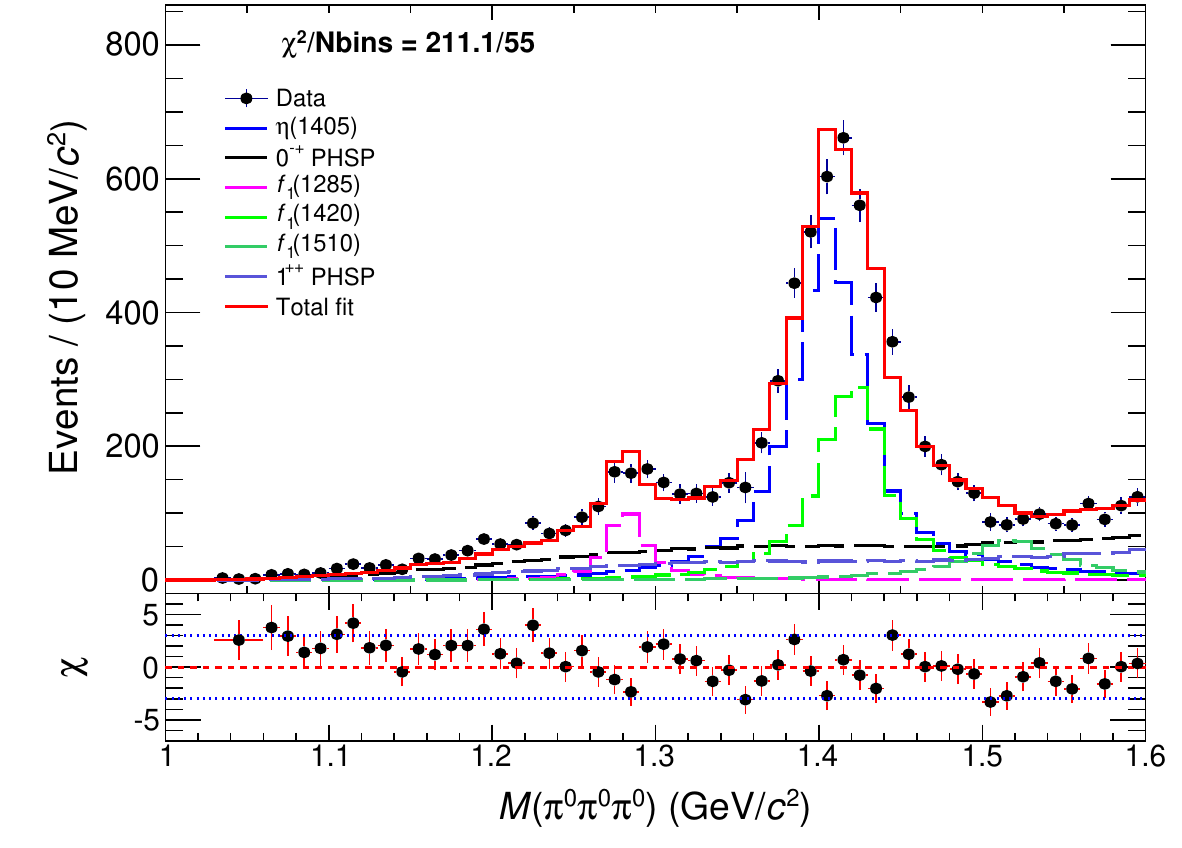} \put(-30,130){(a)} 
	\includegraphics[width=0.38\textwidth]{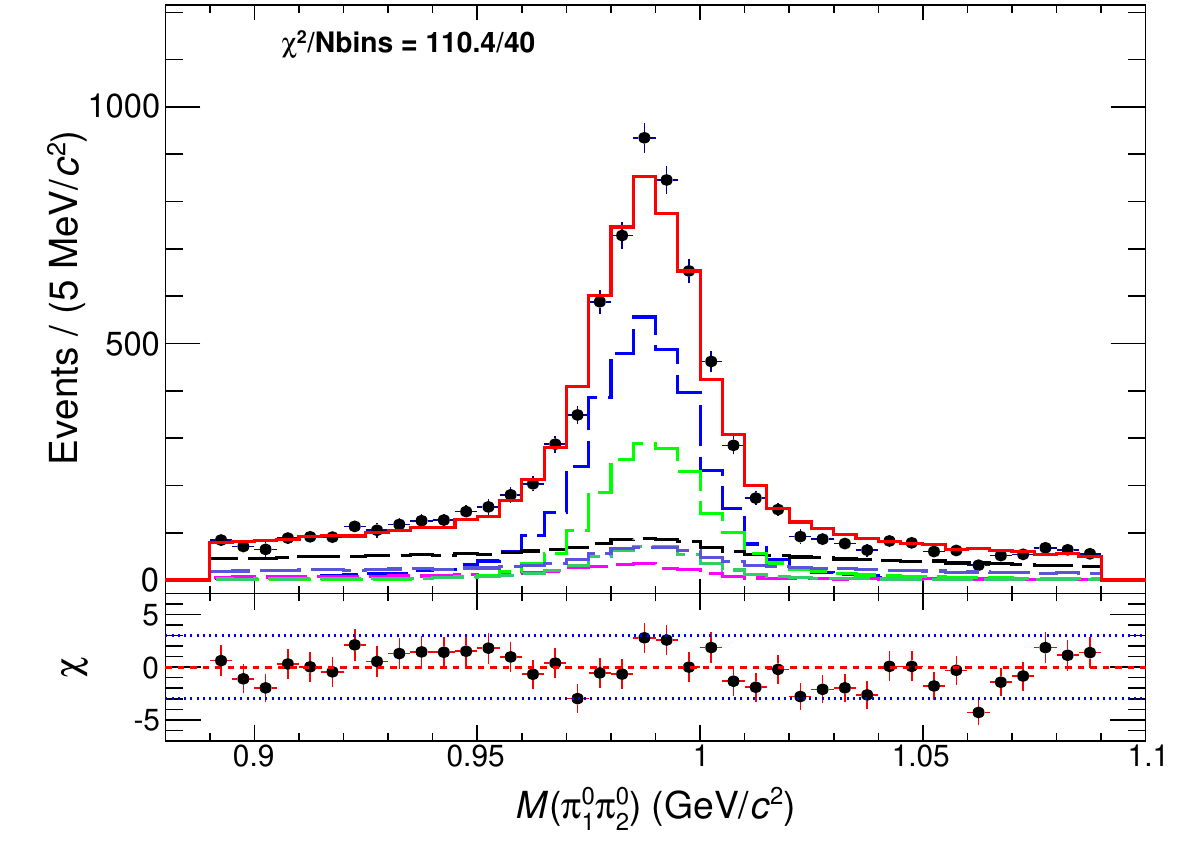} \put(-30,130){(b)} \\
	\includegraphics[width=0.38\textwidth]{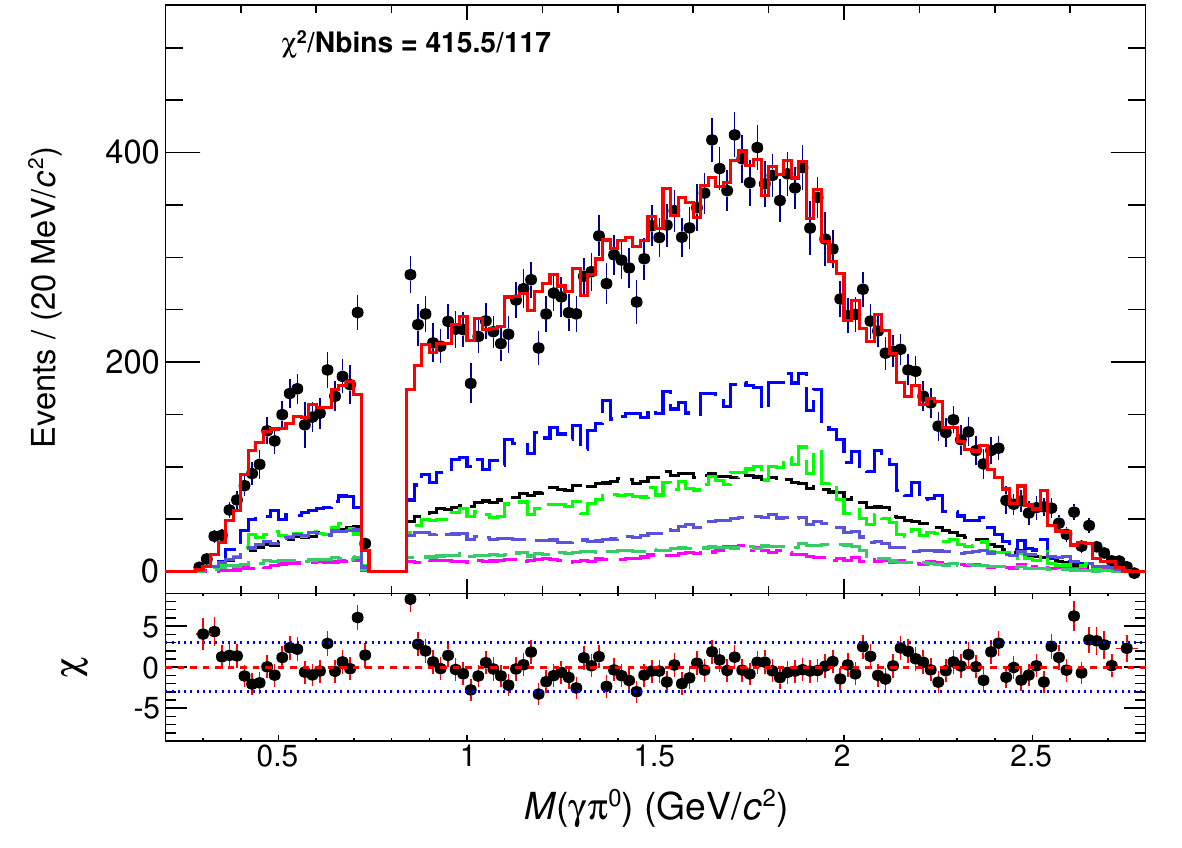} \put(-30,130){(c)}	
	\includegraphics[width=0.38\textwidth]{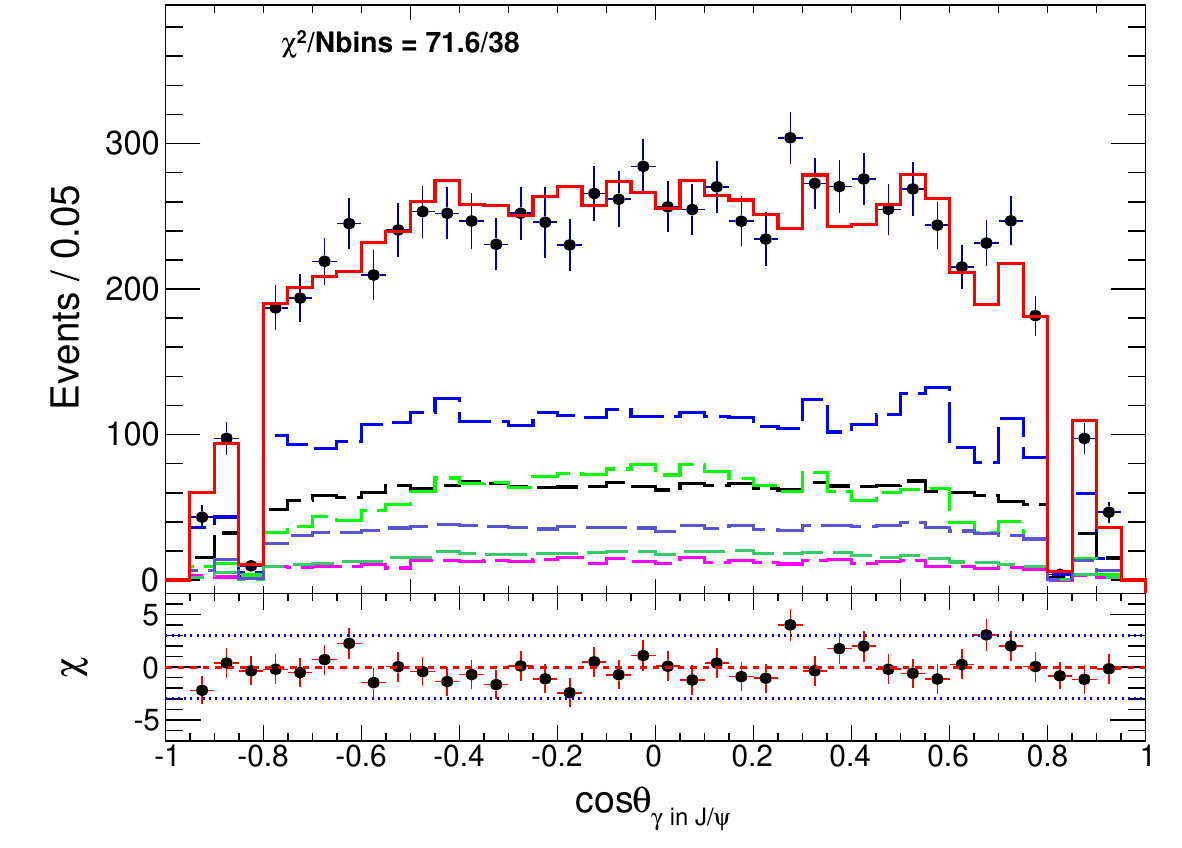} \put(-30,130){(d)}\\ 
	\includegraphics[width=0.38\textwidth]{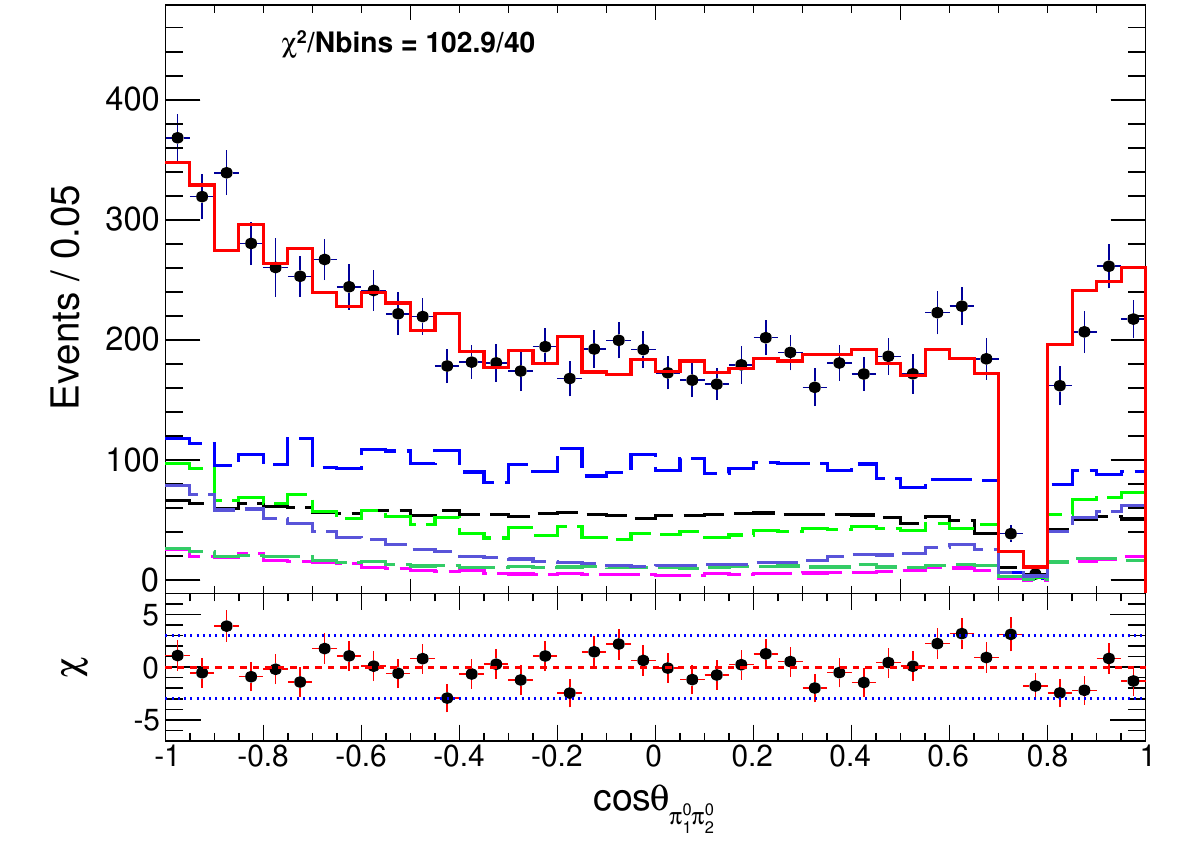} \put(-30,130){(e)}
	\includegraphics[width=0.38\textwidth]{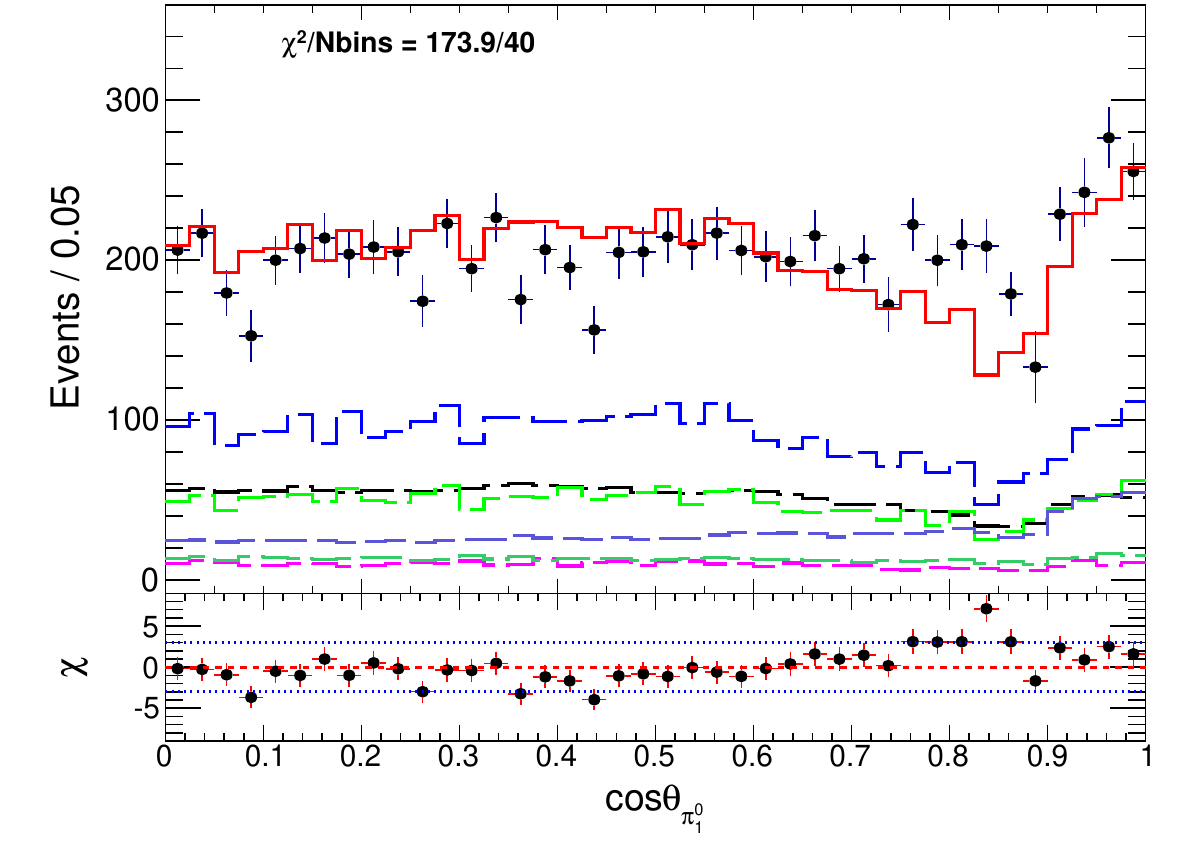} \put(-30,130){(f)} \\ 
	\includegraphics[width=0.38\textwidth]{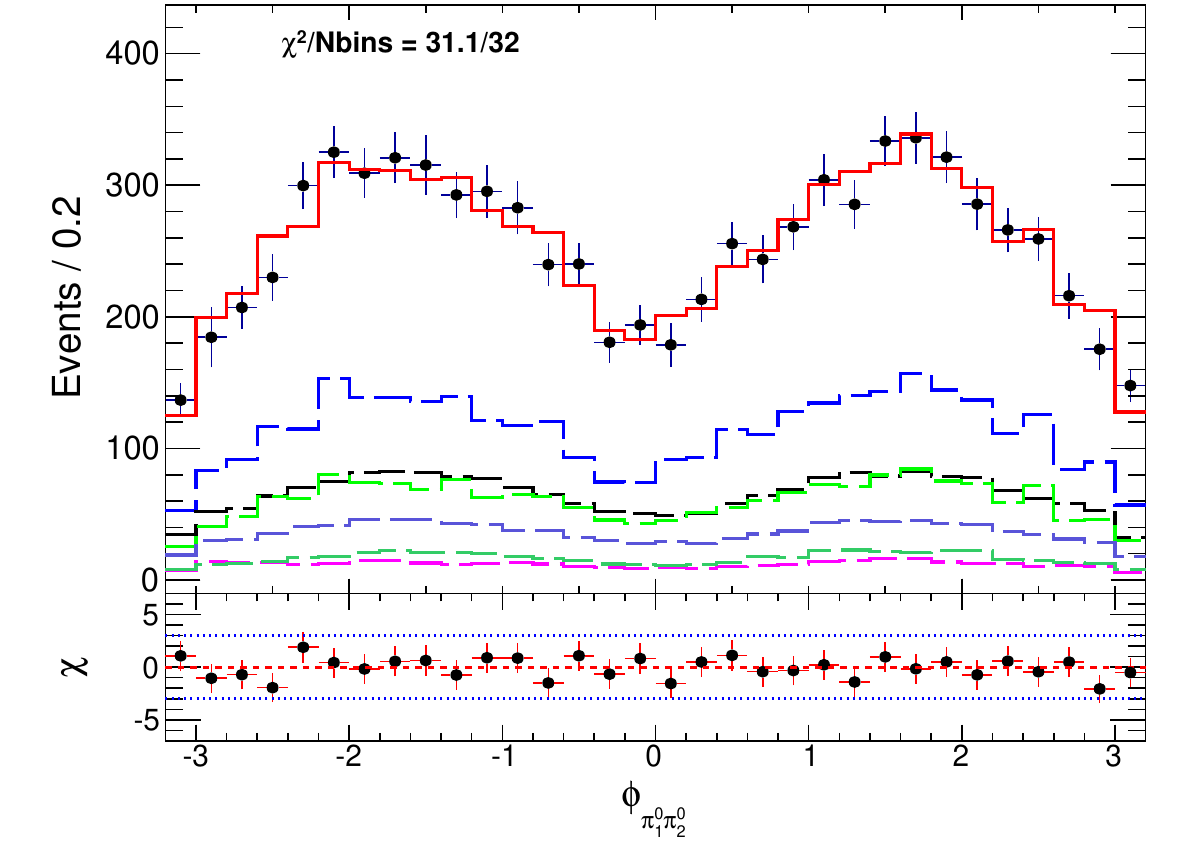} \put(-30,130){(g)}
	\includegraphics[width=0.38\textwidth]{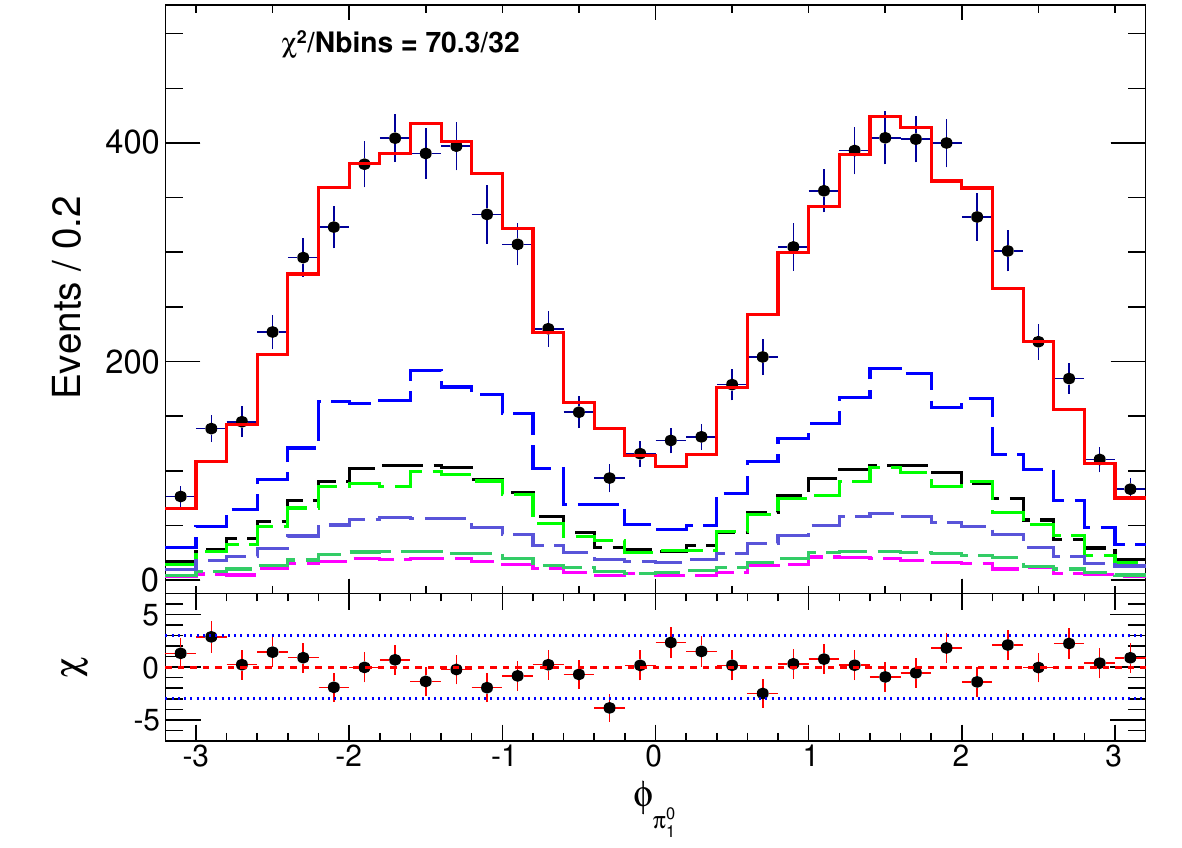} \put(-30,130){(h)}	
	\caption{
	Superposition of data after background subtraction and the mass-dependent PWA fit projections for: 
	the distributions of (a) $M(\tripiz{})$, (b) $M(\piz_1\piz_2)$, where $\piz_1\piz_2$ denotes the closest $\piz\piz$ combination to the mass of $f_0(980)$ where the momentum of $\piz_1$ is larger than that of $\piz_2$, (c) $M(\gam\piz)$; 
	cos$\theta$ of (d) $\gamma$ in the $\jpsi$ rest frame; (e) $\piz_1\piz_2$ in the $\jpsi$ rest frame, (f) $\piz_1$ in the $\piz_1\piz_2$ rest frame; azimuthal angle between the normals to the two decay planes of (g) $\piz_1\piz_2$ in the $\tripiz{}$ rest frame, and (h) $\piz_1$ in the $\piz_1\piz_2$ rest frame. The subplots in the lower panel in each plot are the corresponding pull distributions, where the red and blue lines indicate the reference lines for the center (no deviation) and the deviation of $\pm~3\sigma$, respectively.\label{fig:pwa_fit_result}}
\end{figure*}

\begin{figure*}[htbp!]
	\centering	
	\includegraphics[width=0.50\textwidth]{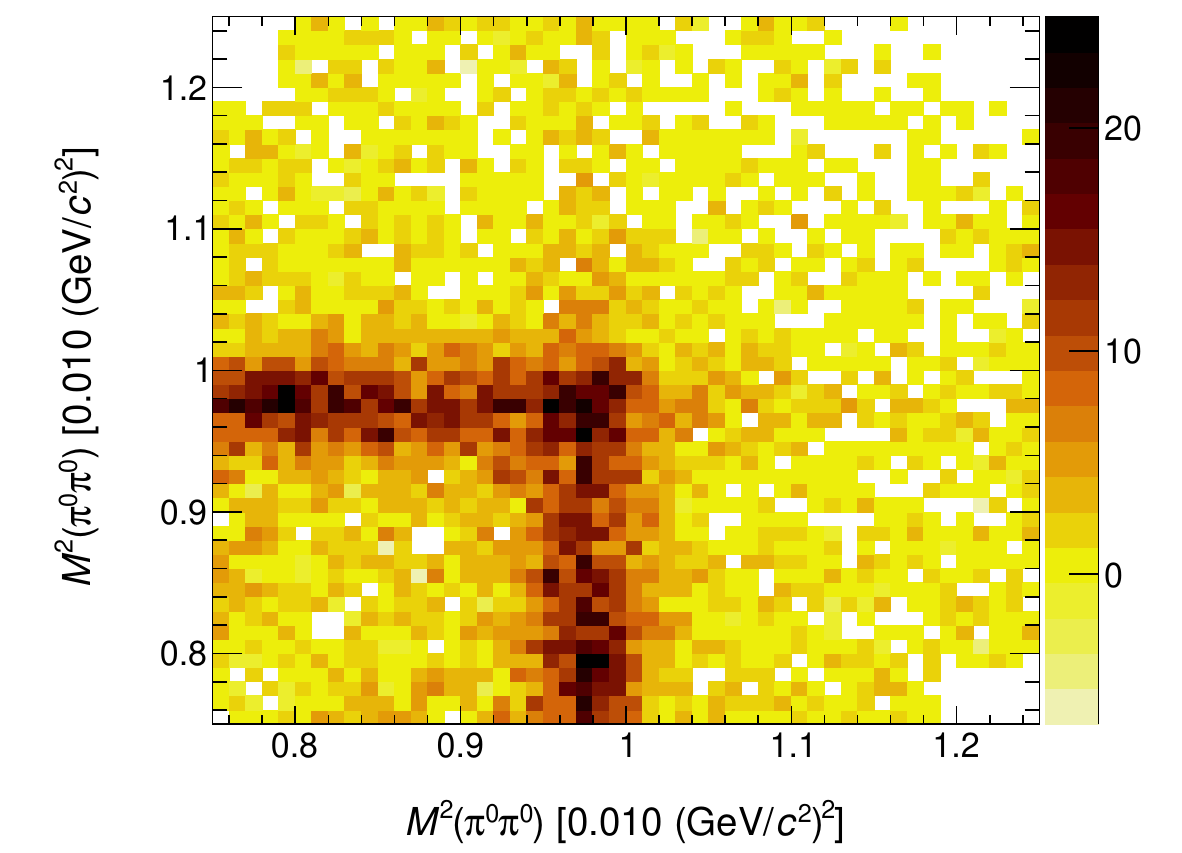} \put(-50,168){(a)} 
	\includegraphics[width=0.50\textwidth]{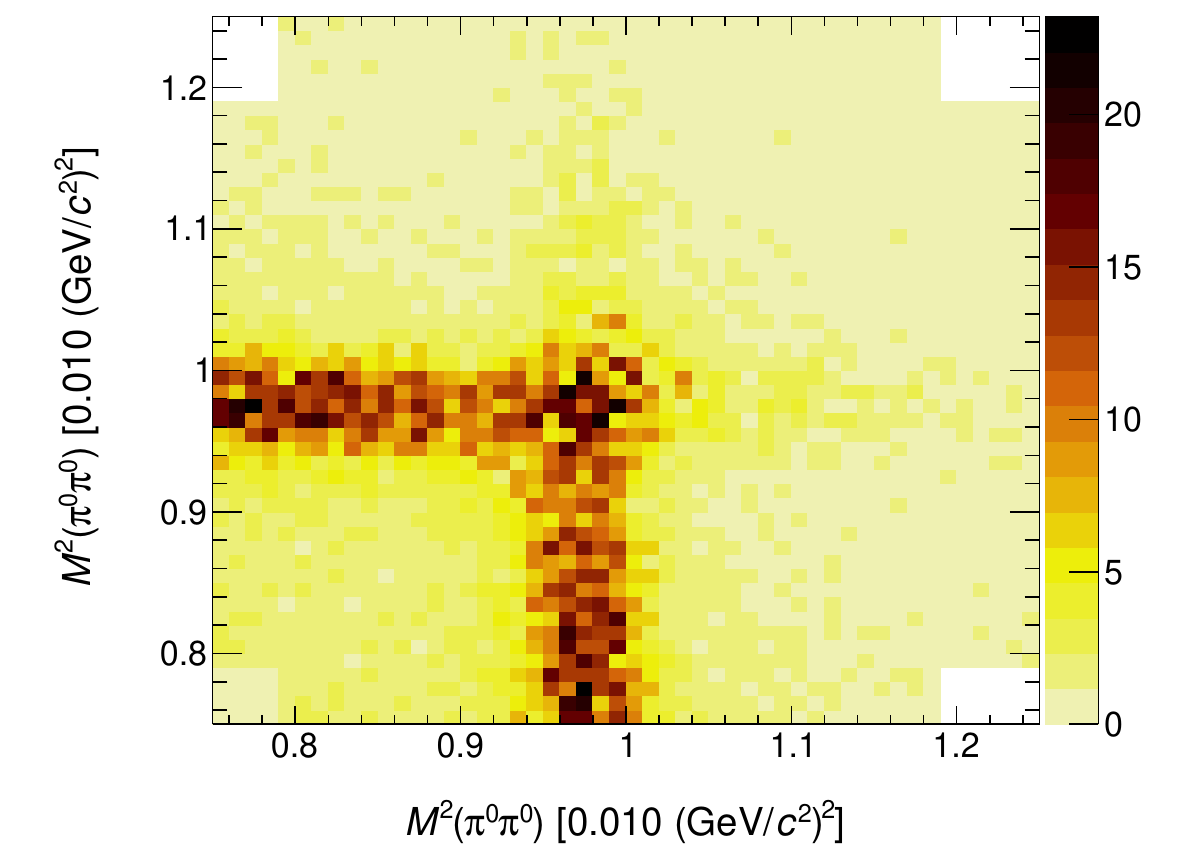} \put(-50,168){(b)} \\
	\caption{
	The Dalitz plots of the $\piz\piz\piz$ system (filled six times for each event) from the (a) data and (b) PWA fit result.\label{fig:pwa_fit_dalitz}}
\end{figure*}

The comparisons of data and the PWA fit projections (weighted by MC efficiencies) of the $M(\tripiz{})$, $M(\piz\piz)$, $M(\gam\piz)$ distributions and the Dalitz plots of the $\piz\piz\piz$ system are shown in Figs.~\ref{fig:pwa_fit_result}(a),~\ref{fig:pwa_fit_result}(b),~\ref{fig:pwa_fit_result}(c) and Fig.~\ref{fig:pwa_fit_dalitz}.
The comparisons of data and the projected MC angular distributions are shown in Figs.~\ref{fig:pwa_fit_result}(d)-(h).
The $\chisq/N_{\rm bins}$ value is displayed on each figure to demonstrate the goodness of fit, where $N_{\rm bins}$ is the number of bins and the $\chisq$ is defined as:
\begin{equation}
	\chisq=\sum^{N_{\rm bins}}_{i=1}\frac{\left(n_i-\nu_i\right)^2}{\nu_i},
\end{equation}
where $n_i$ and $\nu_i$ are the numbers of events in data and the fit projections with the baseline model in the $i$-th bin of each figure, respectively.

\begin{figure*}[htbp!]
	\centering	
	\includegraphics[width=0.45\textwidth]{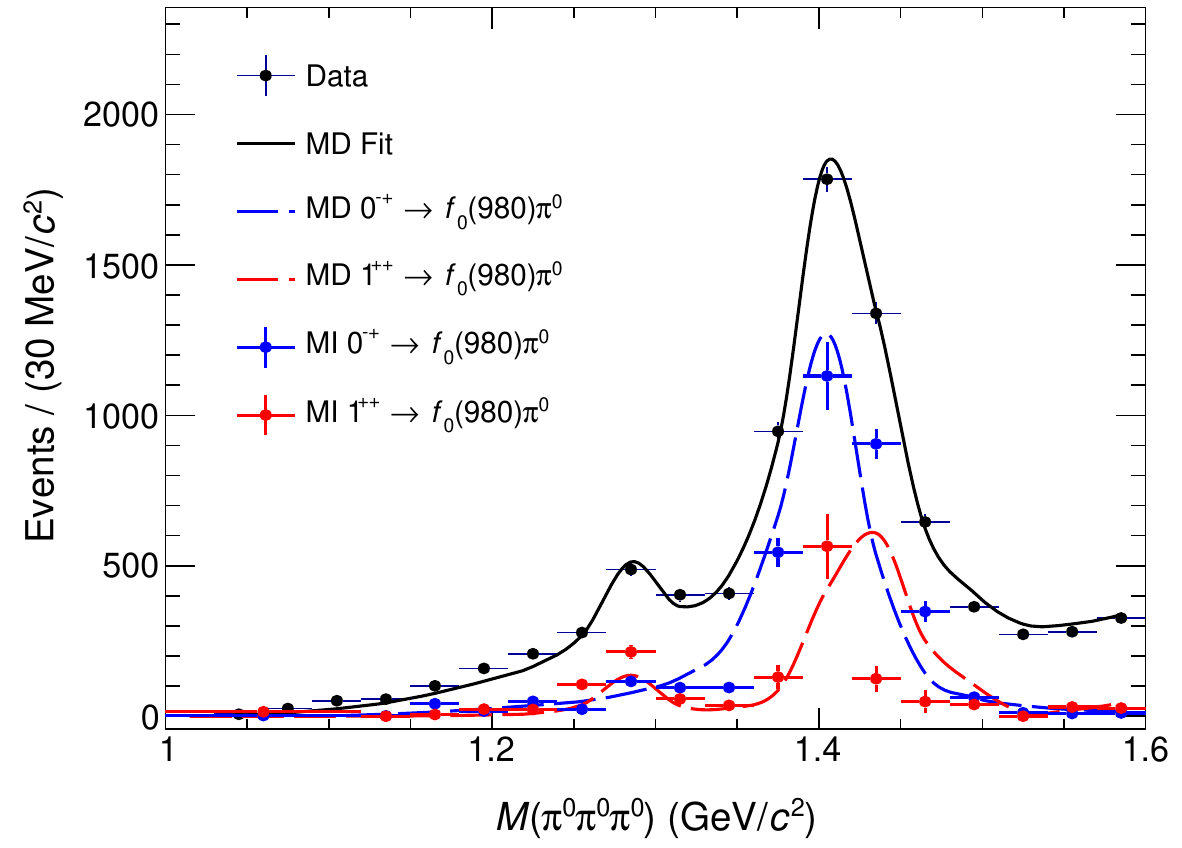} \put(-40,140){(a)}	
	\includegraphics[width=0.45\textwidth]{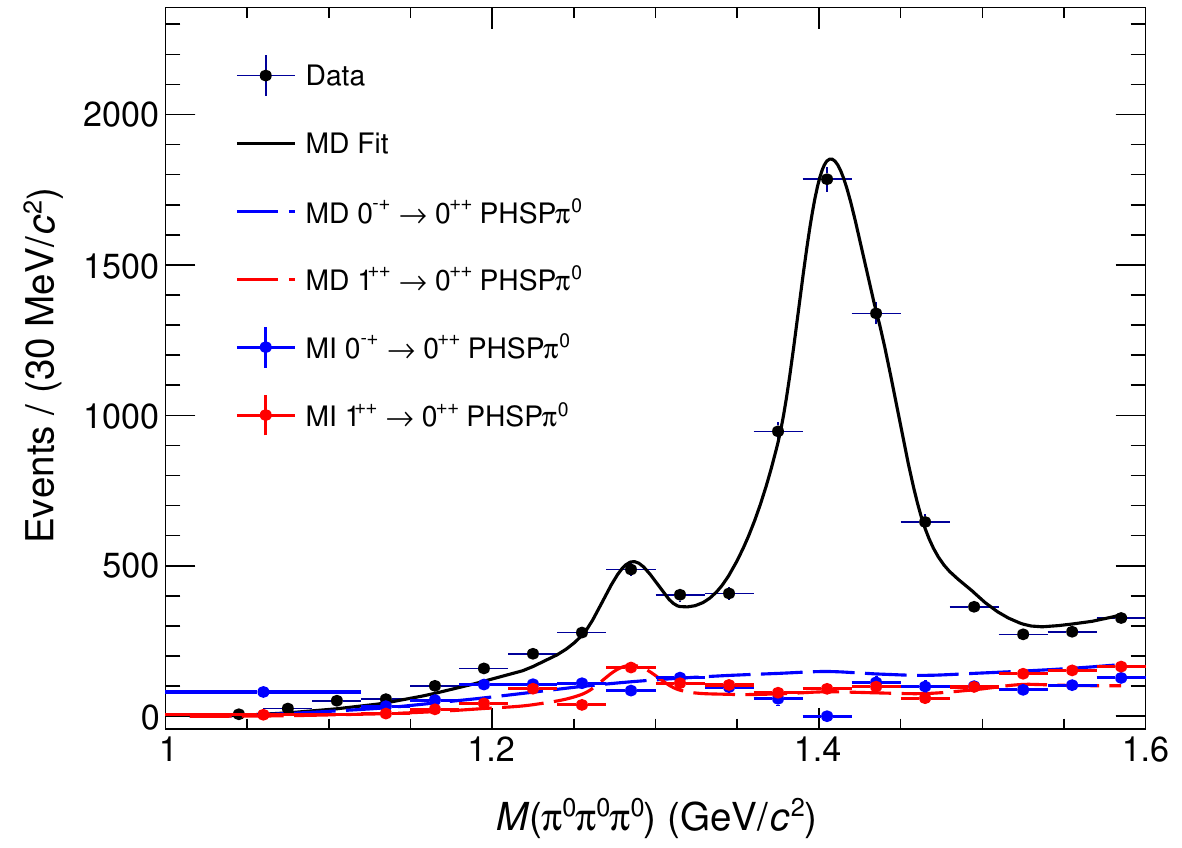} \put(-40,140){(b)}\\
	\vspace*{0.2cm}
	\includegraphics[width=0.45\textwidth]{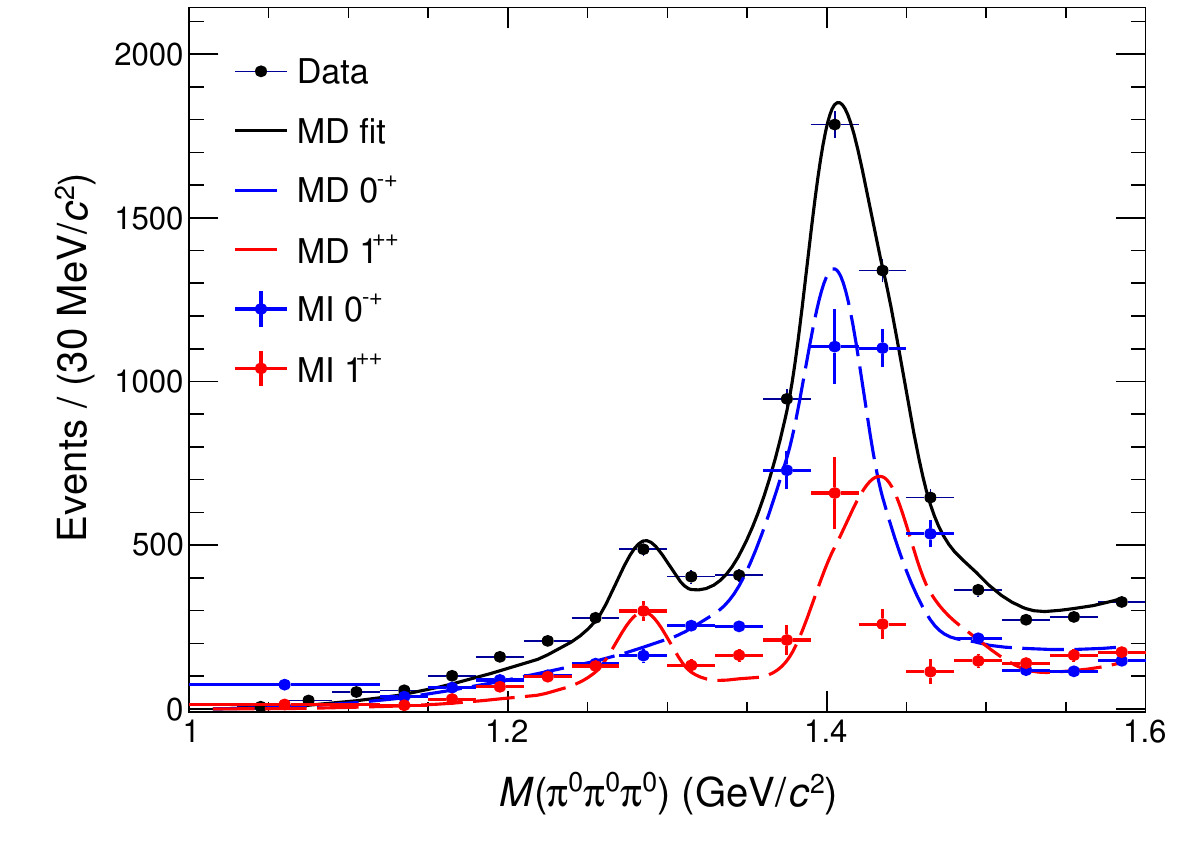} \put(-40,140){(c)}
	\caption{Comparison of the results from the MD PWA and MI PWA fits on $M(\tripiz{})$ for contributions with sub-decays of (a) $f_0(980)\piz$ and (b) $0^{++}$ PHSP$\piz$, and (c) the sub-decays combined.
	In the MI PWA fit, the first four bins are combined due to the low number of events.\label{fig:cp_md_mid}
	}
\end{figure*}

To investigate the contributions from different components, a mass-independent~(MI) PWA is performed as a crosscheck,
which is necessary for studying the line-shapes of different decay processes in the $M(\tripiz{})$ 
distribution and for reducing the biases from a specific model regarding the dynamics of the intermediate states.
Here, we do not consider contributions from the possible processes $\jpsi\to R_1(\to\gam\piz) R_2(\to\piz\piz)$ and $\jpsi\to \piz R(\to\gam\piz\piz)$.
In the MI PWA, the $M(\tripiz{})$ distribution in the range of [1.0, 1.6] GeV/$c^{2}$ is divided into 20 equal bins, 
while the first four bins are combined into one large bin due to the low number of entries.
The intermediate states in the $M(\tripiz{})$ distribution
for each bin are parametrized by an individual complex constant, while the part of the amplitude describing the dynamical function is constant over the small range of invariant-mass squared $s$.
With the MI PWA, the contribution of components for different decay processes can be extracted.
This method has already been applied in Refs.~\cite{binbybin_BES:2003iac,BESIII:2022chl}.
The $0^{-+}$ and $1^{++}$ components on $\tripiz{}$ are also included in our MI PWA.
The comparison of the results from the baseline PWA, i.e. the mass-dependent (MD) PWA,
and the MI PWA fits 
is shown in Fig.~\ref{fig:cp_md_mid}.
The results from the MD PWA and the MI PWA are in qualitative agreement, where the structure around 1.3\,GeV/$c^2$ is dominated by a $1^{++}$ contribution and the structure around 1.4\,GeV/$c^2$ arises from $0^{-+}$ and $1^{++}$ contributions.  However, it is observed that there are discrepancies between the MD PWA and MI PWA  in the line-shapes of the $0^{-+}$ and $1^{++}$ resonances, which could indicated that a more sophisticated model incorporating the coupled-channel effect near the $K^*K$ threshold is required to describe the data.

Various checks are performed based on the PWA baseline model.
If the $0^{++}$ PHSP in $\piz\piz$ is replaced by a $2^{++}$ PHSP component, the NLL value is worsened by 385.8.
The fitted mass and width of the pseudo-scalar 
resonance
around 1.4~GeV/$c^{2}$ is 1404~MeV/$c^2$ and 46~MeV, respectively.
These values are more compatible with the mass and width of the $\eta(1405)$ as given in the PDG~\cite{PDG:2022pth} (1408.8~MeV/$c^2$ and 50.1~MeV) rather than those of the $\eta(1475)$ (1475~MeV/$c^2$ and 90~MeV).
If the mass and width of $\eta(1405)$ in the baseline model are fixed to the PDG resonance parameters of $\eta(1405)$, the NLL value degrades by  9.4.
If the $\eta(1405)$ is replaced with $\eta(1475)$, with resonance parameters fixed to the PDG values, the NLL value increases by 195.6.
Due to the limited sample size and the overlapping of the $\eta(1405)$ and $\eta(1475)$, the interference between them is difficult to resolve.
Thus, the possible existence of $\eta(1475)$ decaying into $\piz\piz\piz$ can not be excluded entirely.
When the $\eta(1295)$ is included in the fit as an additional resonance, its fit fraction is small but with a statistical significance larger than 5$\sigma$.
This can be caused by the possible interference between $\eta(1295)$ and $\eta(1405)$.
In addition, when the $\eta(1295)$ is included, the line-shapes of $0^{-+}$ and $1^{++}$ components of the MD fit exhibit significant discrepancies with the results from the MI fit.
Therefore, $\eta(1295)$ is not included in the baseline solution.
The effects brought by these two states are taken into account when assigning  systematic uncertainties from the presence of possible additional resonances.

\section{Systematic uncertainties} \label{sec:sys}
The common uncertainties include the systematic uncertainties associated with the total number of $\jpsi$ events 
(0.4\%~\cite{numJpsi_BESIII:2021cxx}), photon detection (0.53\% per photon~\cite{photon_detection_BESIII:2017hup,photon_BESIII:2019tel}), kinematic fit (1.2\% from the study of MC), 
and the knowledge of ${\cal B}_{\piz\ra\gam\gam}$ (0.03\% for each $\piz$)~\cite{PDG:2022pth}.
The quadrature sum of them is 3.9\%.
Uncertainties from the PWA fit procedure, where systematic uncertainties are assigned to both the measurements of 
the branching fractions and resonance parameters, has the following contributions.
\begin{itemize}
	\item[(i)] 
	Detector resolution:
	To take into account the systematic uncertainty from the detector resolution on the $f_0(980)$ lineshape in the $M(\piz\piz)$ distribution, 
	the width of the Gaussian function used in the numerical convolution, 9.6~MeV, is increased and decreased by its statistical uncertainty, 0.3~MeV. 
	The maximum change observed with respect to the baseline result is assigned as the corresponding uncertainty.

	\item[(ii)] 
	Additional resonances:

	Uncertainties from possible additional resonances are estimated by adding the $\eta(1295)$ and the $\eta(1475)$, which are the two most significant additional resonances, and also the possible $2^{-+}$ PHSP, into the baseline solution individually. 
	The changes in the results induced by these additional contributions are summed in quadrature and assigned as the systematic uncertainty from the presence of unknown resonances.
	Owing to the close proximity between the $\eta(1405)$ and the $\eta(1475)$, the uncertainty in the branching fraction of $\eta(1405)$ from this source is inherently large.
	Those contributions from resonances at higher masses are negligible. 
	For example, the $\eta(1760)$ cannot be distinguished from $0^{-+}$ PHSP in our fit range.

	\item[(iii)]
	Interferences:
	A PWA fit considering the interferences between $0^{-+}~{\rm PHSP}(\tripiz)\to \piz + 0^{++}~{\rm PHSP}(\piz\piz)$ or $1^{++}~{\rm PHSP}(\tripiz)\to \piz + 0^{++}~{\rm PHSP}(\piz\piz)$ and other processes is performed.
	The difference observed with respect to the baseline result is assigned as the corresponding uncertainty.

	\item[(iv)]
	Background level:
	Alternative fits are performed raising and lowering the background by its uncertainty.  The maximum difference in the result is assigned as the corresponding systematic uncertainty on the branching fractions and resonance parameters.

\end{itemize}
For each alternative fit performed to estimate the systematic uncertainties from the PWA fit procedure, the changes of the results are taken as the one-sided systematic uncertainties.
For each measurement, the individual uncertainties are assumed to be independent and are added in quadrature to obtain the total systematic uncertainty on the negative and positive sides, respectively.
These systematic uncertainties are applied to the measurements of the masses and widths of the $\eta(1405)$ and the $f_1(1420)$, and  are summarized in Table~\ref{tab:un_sum_mw}.
The relative systematic uncertainties on the branching-fraction measurements are summarized in 
Table~\ref{tab:un_sum_br}.
\begin{table*}[htbp]
\small
\centering
\caption{
The systematic uncertainties in the masses (in MeV/$c^{2}$) and widths (in MeV) of $\eta(1405)$ and $f_1(1420)$ in the baseline model, denoted as $\Delta$$M$ and $\Delta\Gamma$.
}
\label{tab:un_sum_mw}
\setlength{\tabcolsep}{2.0 mm}{
	\renewcommand\arraystretch{1.5}
	\begin{tabular} { l c   c   c c}
	\hline \hline 
	\multirow{2}*{Source} & \multicolumn{2}{c }{$\eta(1405)$}  &  \multicolumn{2}{c }{$f_1(1420)$}  \\
	\cline{2-5}
	~ & $\Delta M$ & $\Delta \Gamma $ & $\Delta M$ & $\Delta \Gamma $ \\ \hline
	Detector resolution  &  0.0 & 0.0 & 0.0 & 0.0 \\  
	Extra resonances  & -8.1 & +3.7 & -2.2 & ${}^{+4.5}_{-11.0}$  \\   
	Interferences  & +2.0 & +2.0 & +2.0 & +4.0  \\   
	Background uncertainty & 0.0 & 0.0 & 0.0 & +1.0  \\   
	Total & ${}^{+2.0}_{-8.1}$ & +4.2 & ${}^{+2.0}_{-2.2}$ & ${}^{+6.1}_{-11.0}$ \\ \hline \hline

	\end{tabular} 
	}
\end{table*}

\begin{table*}[htbp!]
\centering
\caption{
The systematic uncertainties (in \%) in the branching-fraction measurements of  the intermediate states of the decay $\jpsi\ra\gam X\ra \gam \tripiz{}$. 
The quadrature sum of common systematic uncertainties (3.9\%) has been added into the last row.
}
\label{tab:un_sum_br}
\setlength{\tabcolsep}{2.0 mm}{
	\renewcommand\arraystretch{1.5}
	\begin{tabular} { l   c   c   c c c c }
	\hline \hline 
	Source & $\eta(1405)$  & $0^{-+}$ PHSP & $f_1(1285)$   & $f_1(1420)$   & $f_1(1510)$   & $1^{++}$ PHSP \\
	\hline
  	Detector resolution  & ${}^{+0.8}_{-0.0}$ & ${}^{+1.2}_{-0.0}$& ${}^{+0.0}_{-2.2}$ & ${}^{+1.4}_{-0.0}$ & ${}^{+0.8}_{-0.0}$ & ${}^{+2.1}_{-0.0}$ \\ 
  	Extra resonances  & ${}^{+104.2}_{-0.0}$ & ${}^{+7.9}_{-46.5}$ & ${}^{+12.6}_{-39.7}$ & ${}^{+7.3}_{-50.7}$ & ${}^{+8.1}_{-19.2}$ & ${}^{+17.9}_{-63.8}$ \\ 
  	Interferences 	 & +34.5 & -4.2 & -36.4 & -17.7 & -44.2 & +53.5 \\ 
  	Background uncertainty	& ${}^{+0.2}_{-0.0}$ & ${}^{+9.2}_{-9.1}$ & ${}^{+0.2}_{-2.8}$ & ${}^{+2.8}_{-0.0}$ & ${}^{+2.6}_{-2.0}$ & ${}^{+6.7}_{-0.0}$  \\ \hline
  	Total & ${}^{+109.8}_{-3.9}$ & ${}^{+12.8}_{-47.7}$ & ${}^{+13.2}_{-54.1}$ & ${}^{+8.8}_{-53.8}$ & ${}^{+9.4}_{-48.4}$ & ${}^{+57.0}_{-63.9}$ \\ \hline \hline

	\end{tabular} 
	}
\end{table*}

\section{Summary}
In summary, a PWA of the decay $\jpsi\ra\gam\piz\piz\piz$ has been performed with $M(\tripiz{})<$ 1.6~GeV$/c^{2}$ based on $(10.09~\pm~0.04)\times10^9~\jpsi$ 
events collected with the BESIII detector.
In contrast to the one-dimensional fit to $M(\tripiz{})$ performed in a previous analysis~\cite{isobreak_jpsitoG3pi_BESIII:2012aa}, 
the current study uses a PWA to disentangle the structures around 1.3 and 1.4 GeV/$c^{2}$ on $M(\tripiz{})$, determining the separate contributions from the 
$f_1(1285)$, $\eta(1405)$, $f_1(1420)$ and $f_1(1510)$ states. 
These results provide valuable inputs for further development of phenomenological model around the $K^{*}\bar{K}$ mass threshold~\cite{tsm_Nakamura:2022rdd,tsm_Nakamura:2023hbt}.

Three axial-vectors, $f_1(1285)$, $f_1(1420)$ and $f_1(1510)$, have been observed in the 
decay to $\tripiz{}$ for the first time, 
providing additional insights in our understanding of the $J^{PC}=1^{++}$ nonet.
The measured production of $f_1(1285)$ in $\jpsi{}\ra\gam\tripiz{}$ is consistent with the upper limit estimation from the previous BESIII study~\cite{isobreak_jpsitoG3pi_BESIII:2012aa}.

The measured production branching fraction of $\eta(1405)$ 
is consistent with the previous study which was obtained from one-dimensional fit to $M(\tripiz{})$~\cite{isobreak_jpsitoG3pi_BESIII:2012aa}.
However, the current results suffer from a large uncertainty due to the possible impact from the $\eta(1475)$.
To reveal the properties of pseudo-scalars around 1.4~GeV$/c^{2}$ and address the puzzle of $\eta(1405)/\eta(1475)$ thoroughly, it requires a sophisticated couple-channel analysis, incorporating the $\ksks{}\piz$ channel~\cite{BESIII:2022chl} and considering triangle singularity mechanism and other phenomenological mechanisms~\cite{tsm_Nakamura:2022rdd,tsm_Nakamura:2023hbt}.

\begin{acknowledgements}
The BESIII Collaboration thanks the staff of BEPCII and the IHEP computing center for their strong support. This work is supported in part by National Key R\&D Program of China under Contracts Nos. 2020YFA0406300, 2020YFA0406400, 2023YFA1606000;
National Natural Sci ence Foundation of China (NSFC) under Contracts Nos. 11635010, 11735014, 11835012, 11922511, 11935015, 11935016, 11935018, 11961141012, 12022510, 12025502, 12035009, 12035013, 12061131003, 12192260, 12192261, 12192262, 12192263, 12192264, 12192265, 12221005, 12225509, 12235017, 12361141819;
the Chinese Academy of Sciences (CAS) Large-Scale Scien tific Facility Program;
the CAS Center for Excellence in Particle Physics (CCEPP);
Joint Large-Scale Scientific Facility Funds of the NSFC and CAS under Contract No. U1832207;
CAS Key Research Program of Frontier Sciences under Contracts Nos. QYZDJ-SSW SLH003, QYZDJ-SSW-SLH040;
100 Talents Program of CAS;
CAS Project for Young Scientists in Basic Research YSBR-101;
The Institute of Nuclear and Particle Physics (INPAC) and Shanghai Key Laboratory for Particle Physics and Cosmology;
ERC under Contract No. 758462;
European Union's Horizon 2020 research and innovation programme under Marie Sklodowska-Curie grant agreement under Contract No. 894790;
German Research Foundation DFG under Contracts Nos. 443159800, 455635585, Collaborative Research Center CRC 1044, FOR5327, GRK 2149;
Istituto Nazionale di Fisica Nucleare, Italy;
Knut and Alice Wallenberg Foundation under Contracts Nos. 2021.0174, 2021.0299; 
Ministry of Development of Turkey under Contract No. DPT2006K-120470;
National Research Foundation of Korea under Contract No. NRF-2022R1A2C1092335;
National Science and Technology fund of Mongolia;
National Science Research and Innovation Fund (NSRF) via the Program Management Unit for Human Resources \& Institutional Development, Research and Innovation of Thailand under Contracts Nos. B16F640076, B50G670107;
Polish National Science Centre under Contract No. 2019/35/O/ST2/02907;
Swedish Research Council under Contract No. 2019.04595;
The Swedish Foundation for International Cooperation in Research and Higher Education under Contract No. CH2018-7756;
U. S. Department of Energy under Contract No. DE-FG02-05ER41374	
\end{acknowledgements}

\bibliography{reference}

\end{document}